\documentclass[10pt]{amsart}
\usepackage[T1]{fontenc}
\usepackage{a4wide}
\usepackage[latin1]{inputenc}
\usepackage{amsmath}
\usepackage{amsfonts}
\usepackage{amssymb}
\usepackage{amsthm}
\usepackage{subfigure}
\newtheorem{theo}{Theorem}[section]
\newtheorem{prop}[theo]{Proposition}
\newtheorem{coro}[theo]{Corollary}
\newtheorem{lemm}[theo]{Lemma}

\theoremstyle{definition}

\theoremstyle{remark}
\newtheorem*{rema}{Remark}
\newcommand{\Op}{\text{Op}}
\newcommand{\U}{\mathcal{U}}

\title{Delocalization of slowly damped eigenmodes on Anosov manifolds}

\author{Gabriel Rivi\`ere}
\address{Laboratoire Paul Painlev\'e (U.M.R. CNRS 8524), U.F.R. de Math\'ematiques, Universit\'e Lille 1, 59655 Villeneuve d'Ascq Cedex, France}
\email{gabriel.riviere@math.univ-lille1.fr}

\begin{document}

\begin{abstract}
We look at the properties of high frequency eigenmodes for the damped wave equation on a compact manifold with an Anosov geodesic flow. We study eigenmodes with spectral parameters which are asymptotically close enough to the real axis. We prove that such modes cannot be completely localized on subsets satisfying a condition of negative topological pressure. As an application, one can deduce the existence of a ``strip'' of logarithmic size without eigenvalues below the real axis under this dynamical assumption on the set of undamped trajectories. 
\end{abstract}

\maketitle

\section{Introduction}

Let $M$ be a smooth, compact riemannian manifold of dimension $d\geq 2$ and without boundary. We will be interested in the high frequency analysis of the damped wave equation,
$$\left(\partial_t^2-\Delta+2V(x)\partial_t\right)u(x,t)=0,\ u(x,0)=u_0,\ \partial_tu(x,0)=u_1,$$
where $\Delta$ is the Laplace-Beltrami operator on $M$ and $V\in\mathcal{C}^{\infty}(M,\mathbb{R}_+)$ is the \emph{damping function}. This problem can be rewritten as
\begin{equation}\label{e:DWE}(-\imath\partial_t+\mathcal{A})\mathbf{u}(t)=0,\end{equation}
where $\mathbf{u}(t):=(u(t),\imath\partial_t u(t))$ and
\begin{equation}\label{e:dampedop}\mathcal{A}=\left(\begin{array}{cc}0&\text{Id}\\-\Delta&-2\imath V\end{array}\right):H^1(M)\times L^2(M)\rightarrow H^1(M)\times L^2(M).\end{equation}

This operator generates a strongly continuous and uniformly bounded semigroup $\mathcal{U}(t)=e^{-\imath t\mathcal{A}}$ on $H^1(M)\times L^2(M)$ which solves~\eqref{e:DWE} -- e.g.~\cite{Hi04}, $\S 2$. Hence, it is quite natural to study the spectral properties of $\mathcal{A}$ in order to understand the behavior of the solutions of~\eqref{e:DWE}. For instance, in~\cite{Leb93}, Lebeau established several important relations (related to the decay of energy of solutions) between the evolution problem~\eqref{e:DWE}, the spectral properties of $\mathcal{A}$ and  the properties of the geodesic flow $(g^t)_t$ on the unit cotangent bundle
$$S^*M:=\left\{(x,\xi)\in T^*M:\|\xi\|_x^2=1\right\}.$$


Recall that the spectrum of this \emph{non selfadjoint} operator is a discrete subset of $\mathbb{C}$ made of countably many eigenvalues $(\tau_n)$ which satisfy $\lim_{n\rightarrow+\infty}\text{Re}\ \tau_n=\pm\infty$. We underline that $\tau$ is an eigenvalue of $\mathcal{A}$ when there exists a non trivial function $u$ in $L^2(M)$ such that
\begin{equation}\label{e:specprob}(-\Delta-\tau^2-2\imath\tau V)u=0.\end{equation}
Hence, each eigenvalue $\tau$ can be associated with a normalized ``eigenmode'' $u$ in $L^2(M)$ which gives raise to the following solution of the damped wave equation
$$v(t,x)=e^{-\imath t\tau}u(x).$$
We also underline that $\text{Im}\ \tau_n\in[-2\|V\|_{\infty},0]$ for every $n$ and that $(\tau,u)$ solves the eigenvalue problem~\eqref{e:specprob} if and only if $(-\overline{\tau},\overline{u})$ solves it~\cite{Non11}. Our main concern in the following will be to \emph{understand the asymptotic properties of slowly damped eigenmodes}. Precisely, we consider sequences $(\tau_n,u_n)_n$ solving~\eqref{e:specprob} with 
$$\text{Re}\ \tau_n\rightarrow+\infty\ \text{and}\ \text{Im}\ \tau_n\rightarrow 0.$$

In the case where $V$ is not identically $0$, Lebeau proved the existence of a constant $C>0$ such that for every $\tau\neq 0$ in the spectrum of $\mathcal{A}$, one has~\cite{Leb93}
\begin{equation}\label{e:Lebaccumulation}
\text{Im}\ \tau\leq-\frac{1}{C}e^{-C|\tau|}.
\end{equation}
Hence, eigenfrequencies cannot accumulate faster than exponentially on the real axis. Moreover, Lebeau also provided in~\cite{Leb93} a geometric situation where this inequality is optimal. An important feature of this example is that it does not satisfy the so-called Geometric Control Condition: 
\begin{equation}\label{e:GCC}
 \exists T_0>0\ \text{such that}\ \forall\rho\in S^*M,\ \{g^t\rho:0\leq t\leq T_0\}\cap\{(x,\xi):V(x)>0\}\neq\emptyset.
\end{equation}
In fact, under this assumption, one can prove that there exists a constant $\gamma>0$ such that for every $\tau\neq 0$ in the spectrum of $\mathcal{A}$, one has~\cite{Leb93, Sj00} 
$$\text{Im}\ \tau\leq-\gamma<0.$$

It is then natural to understand how close to the real axis eigenfrequencies can be under the assumption that the Geometric Control Condition does not hold. For this purpose, we introduce the \emph{set of undamped trajectories}\footnote{By a compactness argument, one can verify that the Geometric Control Condition holds if and only if $\Lambda_V=\emptyset$.} 
\begin{equation}\label{e:rapidconv} 
\Lambda_V=\bigcap_{t\in\mathbb{R}}g^t\{(x,\xi)\in S^*M: V(x)=0\}.
\end{equation}
In fact, even if inequality~\eqref{e:Lebaccumulation} is optimal, there may be some geometric assumptions on $M$ or on $\Lambda_V$ under which the accumulation is much slower than exponential. In recent works, many progresses have been made in understanding the spectral properties of~$\mathcal{A}$ in different geometric situations where $\Lambda_V$ is empty or not -- e.g.~\cite{Sj00, Hi04, BuHi07, Chr07, An10a, Sch10, Sch11, Non11}. We will explain some of these results which are related to ours but before that, we will proceed to a semiclassical reformulation of this spectral problem as in~\cite{Sj00}, $\S  1$.

\subsection*{Semiclassical reformulation} Thanks to the different symmetries of our problem, we will only consider the limit $\text{Re}\ \tau\rightarrow+\infty$. Introduce then $0<\hbar\ll 1$. We will look at eigenfrequencies $\tau$ of order $\hbar^{-1}$ by setting
$$\tau=\frac{\sqrt{2z}}{\hbar},\ \text{where}\ z=\frac{1}{2}+\mathcal{O}(\hbar).$$

With this notation, studying the high frequency eigenmodes of the damped wave equation corresponds to look at sequences $(z(\hbar)=\frac{1}{2}+\mathcal{O}(\hbar))_{0<\hbar\ll 1}$ and $(\psi_{\hbar})_{0<\hbar\ll 1}$ in $L^2(M)$ satisfying\footnote{We underline that for simplicity of exposition, we only deal with operators of this form. However, our approach could be adapted to treat the case of more general families of nonselfadjoint operators like the ones considered in~\cite{Sj00},~$\S 1$}
\begin{equation}\label{e:eigenfunct}
(\mathcal{P}(\hbar,z)-z(\hbar))\psi_{\hbar}=0,\ \text{where}\ \mathcal{P}(\hbar,z):=-\frac{\hbar^2\Delta}{2}-\imath\hbar \sqrt{2z(\hbar)} V(x).
\end{equation}
For every $t$ in $\mathbb{R}$, we also introduce the quantum propagator associated to $\mathcal{P}(\hbar,z)$, i.e.
$$\U_{\hbar}^t:=\exp\left(-\frac{\imath t\mathcal{P}(\hbar,z)}{\hbar}\right).$$


After this semiclassical reduction, the question of the accumulation of the eigenfrequencies to the real axis can be translated in understanding how close to $0$ the \emph{quantum decay rate} $\frac{\text{Im}\ z(\hbar)}{\hbar}$ can be. For $\hbar$ small enough, introduce now

$$\Sigma_{\hbar}=\left\{z(\hbar):\exists \psi_{\hbar}\neq 0\in L^2(M),\ \mathcal{P}(\hbar,z)\psi_{\hbar}=z(\hbar)\psi_{\hbar}\right\}.$$

In~\cite{Sj00}, Sj\"ostrand proved several results on the distribution of this semiclassical spectrum. For instance, he showed that eigenvalues $z(\hbar)$ with $\text{Re}\ z(\hbar)$ in a small box around $1/2$ satisfies a Weyl's law in the semiclassical limit $\hbar\rightarrow0^+.$ Moreover, he proved that, in such boxes, most of the imaginary parts $\frac{\text{Im}\ z(\hbar)}{\hbar}$ concentrate on the ergodic averages of $V$ with respect to the geodesic flow. We refer the reader to~\cite{Sj00} for the precise statements. 

In this semiclassical setting, Lebeau's result reads
$$\exists C>0\ \text{such that for}\ \hbar\ \text{small enough},\ \forall z(\hbar)\in\Sigma_{\hbar},\ \frac{\text{Im}\ z(\hbar)}{\hbar}\leq-\frac{1}{C}e^{-\frac{C}{\hbar}}.$$ 
Under the Geometric Control Condition~\eqref{e:GCC}, one can prove the existence of $\gamma>0$ such that for $\hbar$ small enough, one has $\frac{\text{Im}\ z(\hbar)}{\hbar}\leq-\gamma$~\cite{Leb93, Sj00}: one says that there is a \emph{spectral gap}.

\subsection*{Chaotic dynamics}

It is natural to ask whether the results above can be improved when the manifold $M$ satisfies additional geometric properties. In this article, we will be interested in the specific case where the geodesic flow $(g^t)$ on the unit cotangent bundle $S^*M$ has the Anosov property (manifolds of negative curvature are the main example). This assumption implies that the dynamical system $(S^*M,g^t)$ is \emph{strongly chaotic} (e.g. ergodicity, mixing of the Liouville measure $L$ on $S^*M$). Motivated by the properties of the semiclassical approximation, one can expect to exploit these chaotic dynamical properties to obtain more precise results on the distribution of eigenvalues -- see e.g.~\cite{An10a, Sch10, Sch11, Non11} for applications of this idea. In~\cite{Sch10, Sch11, Non11}, it is proved that, under various assumptions on $V$, there exists a spectral gap below the real axis. We will give the precise statements below. Yet, to our knowledge, the existence of a spectral gap or a rate of convergence of quantum decay rates to zero is not known for a general nontrivial $V$ even in this chaotic setting. 

Our precise aim in this article is to \emph{describe the asymptotic distribution of eigenmodes for which $\frac{\text{Im}\ z(\hbar)}{\hbar}\rightarrow 0$ fast enough when the geodesic flow is chaotic}. We will prove that such eigenmodes must in a certain sense be partly delocalized on $S^*M$. 



\subsection*{Semiclassical measures}

In order to describe the asymptotic properties of these slowly damped eigenmodes, we will use the notion of semiclassical measures~\cite{Bu97, EZ}. Consider a sequence of normalized eigenmodes $(\psi_{\hbar})_{\hbar\rightarrow 0^+}$ satisfying
\begin{equation}\label{e:eigenmode}\mathcal{P}(\hbar,z)\psi_{\hbar}=z(\hbar)\psi_{\hbar},\end{equation}
where $z(\hbar)\rightarrow 1/2$ and $\frac{\text{Im}\ z(\hbar)}{\hbar}\rightarrow 0$ as $\hbar$ tends to $0$. If such modes exist, one must at least have~$\Lambda_V\neq\emptyset.$ For a given sequence $(\psi_{\hbar})_{\hbar\rightarrow 0^+}$, we introduce a family of distributions on the cotangent space $T^*M$, i.e.
\begin{equation}\label{e:distrib}
\forall a\in\mathcal{C}^{\infty}_o(T^*M),\ \mu_{\psi_{\hbar}}(a):=\langle\psi_{\hbar},\Op_{\hbar}(a)\psi_{\hbar}\rangle_{L^2(M)},
\end{equation}
where $\Op_{\hbar}(a)$ is a $\hbar$-pseudodifferential operator (see appendix~\ref{a:pdo}). This distribution tells us where the eigenfunction $\psi_{\hbar}$ is located 
on the phase space $T^*M$ and one can try to describe the accumulation points of this sequence of distributions in order to understand the asymptotic localization of $\psi_{\hbar}$. Using classical results from semiclassical analysis~\cite{EZ} (Chapter $5$), one can verify that any accumulation point $\mu$ (as $\hbar\rightarrow0$) of the sequence 
$(\mu_{\psi_{\hbar}})_{\hbar}$ is a probability measure with support in the unit cotangent bundle $S^*M$ and which is invariant under the geodesic flow\footnote{The fact that the measure is exactly invariant under the geodesic flow relies on the fact that $\frac{\text{Im}\ z(\hbar)}{\hbar}\rightarrow 0$ as $\hbar$ tends to $0$. In the case $V=0$, this condition is obviously satisfied.}~$g^t$.


We will call \emph{semiclassical measure} any accumulation point $\mu$ (as $\hbar$ tends to $0$) of a sequence of the form $(\mu_{\psi_{\hbar}})$, where $\psi_{\hbar}$ 
satisfies equation~(\ref{e:eigenfunct}). We will denote
$$\mathcal{M}\left((\psi_{\hbar})_{\hbar\rightarrow 0^+}\right)$$
this set of semiclassical measures associated to the sequence $(\psi_{\hbar})_{\hbar\rightarrow 0^+}$. Under our assumption, it forms a subset of the set $\mathcal{M}(S^*M, g^t)$ of $g^t$-invariant probability 
measure on $S^*M$. Hence, $\mathcal{M}\left((\psi_{\hbar})_{\hbar\rightarrow 0^+}\right)$ is a subset of a natural family in ergodic theory and our precise goal is then to give ergodic properties on its elements. Finally, one can verify that any the support of any $\mu$ in $\mathcal{M}\left((\psi_{\hbar})_{\hbar\rightarrow 0^+}\right)$ is included in the \emph{weakly undamped set} 

$$\mathcal{N}_V:=\overline{\bigcup_{\mu\in\mathcal{M}(S^*M,g^t)}\left\{\text{supp}(\mu):\mu(V)=0\right\}},$$
which is a subset of $\Lambda_V$. It is explained in~\cite{Sch11} that $\Lambda_V$ and $\mathcal{N}_V$ could be different; yet, we would like to mention that, in our setting, the Geometric Condition Condition~\eqref{e:GCC} is also equivalent to $\mathcal{N}_V=\emptyset$.

\section{Main results}

Motivated by questions concerning the Quantum Unique Ergodicity Conjecture\footnote{We refer the reader to~\cite{Non09, Sa11, Ze09} for recent reviews on these questions.}, Anantharaman studied the Kolmogorov-Sinai entropy of semiclassical measures in the case of eigenfunctions of the Laplacian on Anosov manifolds~\cite{An08} -- see also~\cite{BouLin03, Lin06} for earlier results due to Bourgain and Lindenstrauss in an arithmetic setting. Precisely, her result concerns the selfadjoint case $V\equiv0$. It roughly says that, in this context, eigenmodes must be partly delocalized on $S^*M$ (for instance, they cannot concentrate only on closed geodesics). In this article, we will prove similar results in the non selfadjoint setting $V\geq 0$.

Before giving details, we would like to recall that the Kolmogorov-Sinai entropy $h_{KS}(\mu,g)$ 
is a nonnegative quantity associated to an invariant probability $\mu$ in $\mathcal{M}(S^*M,g^t)$ -- see~\cite{Wa} or section~\ref{s:background} for a brief reminder). This quantity characterizes what the 
measure perceives of the complexity of the geodesic flow. For instance, if $\mu$ is carried by a closed orbit of the geodesic flow, then $h_{KS}(\mu,g)=0$. On the other 
hand, if $\mu=L$, the measure has a ``good understanding'' of the complexity of the dynamic and so it has a large entropy. Moreover, entropy is affine with respect to the ergodic decomposition of a measure $\mu$~\cite{EinLin06}. In fact, recall that thanks to the Birkhoff Ergodic Theorem, one knows that, for $\mu$ almost every $\rho$ in $S^*M$,
$$\frac{1}{T}\int_0^T\delta_{g^s\rho}ds\rightharpoonup \mu_{\rho},\ \text{as}\ T\rightarrow+\infty,$$
where $\delta_w$ is the Dirac measure in $w\in S^*M$. The measure $\mu_{\rho}$ is ergodic and one has the ergodic decomposition
$\mu=\int_{S^*M}\mu_{\rho}d\mu(\rho)$. Then, the Kolmogorov-Sinai entropy satisfies
\begin{equation}\label{e:affine}
h_{KS}(\mu,g)=\int_{S^*M}h_{KS}(\mu_{\rho},g)d\mu(\rho).
\end{equation}

\subsection{Main result}
We can now state our main result which is the following:


\begin{theo}\label{t:maintheo1} Suppose $(S^*M,g^t)$ satisfies the Anosov property. Let $P_0$ be a positive constant. There exists $c(P_0)>0$ and $C(P_0)>0$ depending only on $P_0$, on $V$ and on $M$ such that if
\begin{itemize}
\item $(\psi_{\hbar})_{\hbar\rightarrow 0^+}$ is a sequence of eigenmodes satisfying~(\ref{e:eigenmode}) with
$$\forall 0<\hbar\leq \hbar_0,\ z(\hbar)\in\left[\frac{1}{2}-\hbar, \frac{1}{2}+\hbar\right] +\imath\left[ -C(P_0)\frac{\hbar}{|\log\hbar|},+\infty\right];$$
\item $\mu$ is in 
$\mathcal{M}\left((\psi_{\hbar})_{\hbar\rightarrow 0^+}\right)$,
\end{itemize}
then, one has
$$\mu\left(\left\{\rho\in S^*M:h_{KS}(\mu_{\rho},g)\geq-\frac{1}{2}\int_{S^*M}\log J^ud\mu_{\rho}-P_0\right\}\right)\geq
c(P_0),$$
where $J^u(\rho)$ is the unstable Jacobian, i.e. $\displaystyle J^u(\rho):=\left|\det\left(d_{g^1\rho}g^{-1}_{|E^u(g^1\rho)|}\right)\right|$. 
\end{theo}

We will recall in section~\ref{s:background} basic facts on entropy and Anosov systems (in particular the definition of $J^u$~\cite{KaHa}). We underline that, for any invariant probability measure $\nu\in\mathcal{M}(S^*M,g^t)$, the quantity $-\int_{S^*M}\log J^ud\nu$ is positive.

\subsection{Comments}

The lower bound that appears in our theorem is a natural dynamical quantity. In fact, for any invariant probability measure $\nu$ in 
$\mathcal{M}(S^*M,g^t)$, one has the Ruelle-Margulis upper bound~\cite{Rue78}, i.e.
\begin{equation}\label{e:Ruelle}
h_{KS}(\nu,g)\leq-\int_{S^*M}\log J^ud\nu,
\end{equation}
with equality if and only if $\nu$ is the Liouville measure on $S^*M$~\cite{LY85}.  Thus, if one has $\frac{\text{Im} (z(\hbar))}{\hbar}=o\left(|\log\hbar|^{-1}\right)$, our result states that a semiclassical measure of our problem must have ergodic components which are close to be ``half delocalized''.

In the case $V\equiv 0$, Anantharaman proved that this theorem holds~\cite{An08} if we replace the quantity $-\int_{S^*M}\log J^ud\mu_{\rho}$ by 
$$\Lambda_{\min}=\inf_{\nu\in\mathcal{M}(S^*M,g^t)}\left\{-\int_{S^*M}\log J^ud\nu\right\}>0.$$ 
In particular, our result improves this earlier result of Anantharaman in the selfadjoint case $V\equiv 0$. Yet, our main interest here was to show that these entropic properties remain true for slowly damped eigenmodes in the high frequency limit of~\eqref{e:DWE}. In particular, our result shows that semiclassical measures of such modes cannot be carried only by closed orbits of the geodesic flow. 

We underline that Anantharaman proved her result in the general setting of quasimodes satisfying $\|(-\hbar^2\Delta-1)\psi_{\hbar}\|=\mathcal{O}(\hbar/|\log\hbar|)$. In our setting, the sequences of states satisfy a priori only $\|(-\hbar^2\Delta-1)\psi_{\hbar}\|=\mathcal{O}(\hbar)$. For simplicity of exposition, we only treat the case of eigenmodes satisfying~\eqref{e:eigenmode} (in principle, the case of quasimodes could also be derived combining our inputs to the strategy in~\cite{An08}). An interesting extension of our main theorem would also be to understand if the entropic results in~\cite{AN07, Riv10a} can also be adapted in this non selfadjoint situtation.

Our assumption on the rate of convergence of $\frac{\text{Im} (z(\hbar))}{\hbar}$ is a very strong assumption. Except in the case $V\equiv 0$, it is not clear if it can be satisfied by a sequence of eigenvalues. In a subsequent work~\cite{Riv11b}, we will describe (weaker) properties that can be derived in the case where we only suppose $\frac{\text{Im} (z(\hbar))}{\hbar}\rightarrow\beta$.

However, let us mention an interesting consequence of our main Theorem:
\begin{coro}\label{c:loggap} Suppose $(S^*M,g^t)$ satisfies the Anosov property. Let $V\geq 0$ be a smooth function on $M$ such that $ \mathcal{N}_V$ is a nonempty subset satisfying
\begin{equation}\label{e:negpressure}
P_{top}\left( \mathcal{N}_V,g^t,\frac{1}{2}\log J^u\right)<0,
\end{equation}
where $\displaystyle P_{top}\left( \mathcal{N}_V,g^t,\frac{1}{2}\log J^u\right)$ is the topological pressure of $\mathcal{N}_V$ with respect to $\frac{1}{2}\log J^u$. 

Then, there exists a positive constant $C$ and $\hbar_{C}>0$ such that for every $0<\hbar\leq\hbar_C$,
$$\Sigma_{\hbar}\cap \left(\left[\frac{1}{2}-\hbar, \frac{1}{2}+\hbar\right] +\imath\left[ -\frac{C\hbar}{|\log\hbar|},+\infty\right]\right)=\emptyset.$$

\end{coro}

This result establishes the presence of a logarithmic strip without eigenvalues in the case where $\mathcal{N}_V$ satisfies a condition of negative topological pressure. In the case of \emph{surfaces}, such a condition holds if $\mathcal{N}_V$ has Hausdorff dimension $<2$~\cite{BaWo07}. This condition on the topological pressure already appeared in~\cite{NZ09} where it was used to establish a spectral gap for resonances. It was also used in previous works concerning the damped wave equation on Anosov compact manifolds~\cite{Sch10, Sch11, Non11}. 

Based on his work~\cite{Sch10}, Schenck proved in~\cite{Sch11} that under condition~\eqref{e:negpressure}, one can find $\alpha>0$ large enough such that there is a \emph{spectral gap} for the problem associated to the damping function $\alpha V$. We underline that we do not make any assumption on the amplitude of $V$ in our corollary but we only obtain a ``logarithmic'' spectral gap. Recently, Nonnenmacher announced that he can remove Schenck's assumption on the amplitude of $V$ if $M$ is of \emph{constant negative curvature}~\cite{Non11}. Hence, in this specific setting, his result is stronger than ours for any damping $V$. In variable curvature, he obtains a spectral gap under a geometric condition which is less general than~\eqref{e:negpressure} (especially if the Lyapunov exponents fluctuate a lot). Compared with these different works, corollary~\ref{c:loggap} allows only to derive a ``logarithmic'' spectral gap which is probably suboptimal -- see section $3.4$ in~\cite{Non11}. Yet, it has the advantage to hold for any damping (even if it is of low amplitude) and for more general geometric situtations.

As a final comment, we would like to mention the results of Christianson in~\cite{Chr07} which have similarities with corollary~\ref{c:loggap}. The main ``dynamical'' differences are that he does not make any assumption on the global structure of the geodesic flow but that he only considers the case where $\mathcal{N}_V$ is a closed hyperbolic geodesic. It would suggest that corollary~\ref{c:loggap} should be true without any assumption on the global structure of the geodesic flow.


\subsection{Some words about the proof}\label{pa:words}

The general strategy of our proof follows the one from~\cite{An08}. Without getting into the details of the proof, we would like to mention what are the main differences which allows the improvements presented above. In this reference, the general strategy was to use \emph{hyperbolic dispersive estimates}~\eqref{e:hypest} in order to prove that eigenmodes cannot concentrate entirely on subsets of small topological entropy (meaning $<\frac{\Lambda_{\min}}{2}$). Regarding the estimates~\eqref{e:hypest}, it was natural to expect that the results of Anantharaman could be improved using ``thermodynamical quantities'' like topological pressure. This allows to take more into account the variations of the unstable jacobian $J^u$ and to improve slightly the entropic lower bounds from~\cite{An08}. Our first input is to introduce these quantities and to prove that eigenmodes cannot concentrate entirely on subsets of small topological pressure with respect to $\frac{1}{2}\log J^u$ (meaning $<0$). Translated in a discrete setting, it is exactly the statement of proposition~\ref{p:crucialprop}.

An additional difficulty we have to face here is that we want to extend the results to the eigenmodes of a non selfadjoint operator. We would like now to illustrate the kind of difficulties created by this generalization. Thanks to the long time Egorov property~\cite{BoRo02} (see also paragraph~\ref{pa:egorov}), one can verify that, for every $a$ in $\mathcal{C}^{\infty}_c(T^*M)$, there exists $\kappa>0$ such that
$$\forall\ 0\leq t\leq\kappa|\log\hbar|,\ \mu_{\psi_{\hbar}}(a)=e^{-\frac{2t\text{Im}\ z(\hbar)}{\hbar}}\left\langle\psi_{\hbar},\Op_{\hbar}\left(a\circ g^t e^{-2\int_0^t V\circ g^sds}\right)\psi_{\hbar}\right\rangle+\mathcal{O}(\hbar^{\nu}),$$
where $\nu>0$ and the constant in the remainder is uniform for $0\leq t\leq\kappa|\log\hbar|$. Hence, in the case $V\equiv 0$, the distibution $\mu_{\psi_{\hbar}}$ is invariant\footnote{In the case of quasi modes~\cite{An08}, this property remains true but with a worst remainder term.} under $g^t$ modulo small error terms which are uniform for logarithmic times in $\hbar$. This invariance property for logarithmic times was extensively used in~\cite{An08} under various forms. In the case where $V$ is non trivial, we did not find a simple equivalent of this ``pseudo-invariance'' property for long times in $\hbar$. However, instead of it, we make a simple observation that we will use at different steps of our proof -- e.g. proposition~\ref{p:subinvariance}. In fact, one can remark that the quantization procedure is ``almost positive'' (see paragraph~\ref{p:antiwick}); hence, as $V\geq 0$, there exists, for any $a$ in $\mathcal{C}^{\infty}_c(T^*M,\mathbb{R}_+)$, $\kappa>0$ such that
$$\forall\ 0\leq t\leq\kappa|\log\hbar|,\ \mu_{\psi_{\hbar}}(a)\leq e^{-\frac{2t\text{Im}\ z(\hbar)}{\hbar}}\mu_{\psi_{\hbar}}(a\circ g^t)+\mathcal{O}(\hbar^{\nu}).$$
Under the assumption that $\frac{\text{Im}\ z(\hbar)}{\hbar}\geq-C|\log\hbar|^{-1}$, one can then verify that the distribution
satisfies $\mu_{\psi_{\hbar}}(a)\leq e^{2C\kappa}\mu_{\psi_{\hbar}}(a\circ g^t)+\mathcal{O}(\hbar^{\nu})$ with an uniform remainder for $0\leq t\leq\kappa|\log\hbar|$. Moreover, one can choose $C$ small enough to have $e^{2C\kappa}$ arbitrarly close to $1$. In this sense, the distribution $\mu_{\psi_{\hbar}}$ is subinvariant under the geodesic flow for logarithmic times (modulo small error terms) and this kind of property will be sufficient to prove our result.

\subsection{Organization of the article}

In section~\ref{s:background}, we give a brief reminder on the dynamical systems concepts we will use in this article. In section~\ref{s:symboliccoding}, we proceed to a discretization of the manifold which allows to give a symbolic interpretation of the quantum system. The main result of this section is proposition~\ref{p:crucialprop} which shows that eigenmodes cannot concentrate entirely on subsets of small topological pressure. In section~\ref{s:proofs}, we use this result to derive Theorem~\ref{t:maintheo1}. Then, in section~\ref{s:intermediary}, we give the proof of several lemmas that we used to prove proposition~\ref{p:crucialprop}. Finally, in the appendix, we give several results on semiclassical analysis related to our problem and that we used at different stages of the proof.

\section{Background on dynamical systems}\label{s:background}

In this section, we draw a short review on Anosov flows and thermodynamical formalism. We refer the reader to the classical references on this subject for more details, e.g.~\cite{KaHa, Wa}.

\subsection{Anosov flows}

In all this article, we make the assumption that the geodesic flow satisfies the Anosov property. It means that, for every $E>0$ and for every 
$$\rho\in H_0^{-1}\left(\{E\}\right):=\left\{(x,\xi)\in T^*M:H_0(x,\xi)=\frac{\|\xi\|_x^2}{2}=E\right\},$$ 
one has the following decomposition~\cite{KaHa}
$$T_{\rho}H_0^{-1}\left(\{E\}\right)=\mathbb{R}X_p(\rho)\oplus E^u(\rho)\oplus E^s(\rho).$$
In the previous decomposition, $\mathbb{R}X_p(\rho)$ is the direction of the Hamiltonian vector field, $E^u(\rho)$ is the unstable space and $E^s(\rho)$ is the stable space. For every $E>0$, there exists a constant $C>0$ and $0<\lambda<1$ such that for every $t\geq 0$, one has
$$\forall v^u\in E^u(\rho),\ \|d_{\rho}g^{-t}v^u\|\leq C\lambda^{t}\|v^u\|\ \text{and}\ \forall v^s\in E^s(\rho),\ \|d_{\rho}g^{t}v^s\|\leq C\lambda^{t}\|v^s\|.$$

Define now the unstable Jacobian at point $\rho\in S^*M$ and time $t\geq 0$
$$J^u_t(\rho):=\left|\det\left(d_{g^t\rho}g^{-t}_{|E^u(g^t\rho)}\right)\right|,$$
where the unstable spaces at $\rho$ and $g^t\rho$ are equipped with the induced riemannian metric. This defines an H\"older continuous function on $S^*M$~\cite{KaHa} (that can be extended to any energy layer $p^{-1}\left(\{E\}\right)$). We underline that this quantity tends to $0$ as $t$ tends to infinity at an exponential rate. Moreover, it satisfies the following multiplicative property
$$J^u_{t+t'}(\rho)=J^u_t(g^{t'}\rho)J^u_{t'}(\rho).$$
In the following, we will use the notation $J^u(\rho)=J^u_{1}(\rho).$ Finally, we underline that these unstable Jacobian are related to the Lyapunov exponents~\cite{KaHa}, in the sense that, for a given $\mu\in\mathcal{M}(S^*M,g^t)$, one has, for $\mu$ almost every $\rho$,
$$\lim_{t\rightarrow+\infty}-\frac{1}{t}\log J^u_t(\rho)=\sum_{j=1}^{d-1}\chi_j^+(\rho),$$
where the $\chi_j^+(\rho)$ are the positive Lyapunov exponents at point $\rho$. We underline that this last quantity is also equal to $-\int_{S^*M}\log J^u d\mu_{\rho}$.

\subsection{Kolmogorov Sinai entropy} There are several ways to define Kolmogorov-Sinai entropy and we refer the reader to~\cite{Wa} (chapter $4$) for the classical definition and the fundamental properties of entropy. This quantity associate to a $g^t$-invariant measure $\mu$ a nonnegative number that characterize the \emph{complexity of the geodesic flow from the point of view of $\mu$}. A way to define it is to start from a partition $\mathcal{P}=(P_i)_{i=1}^K$ of $S^*M$. Then, for every $\rho$ in $S^*M$ and for every $n$ in $\mathbb{N}$, there exists an unique sequence $(\alpha_0,\ldots,\alpha_{n-1})$ in $\{1,\ldots, K\}^n$ such that $\rho$ belongs to $P_{\alpha_0}\cap g^{-1}P_{\alpha_1}\ldots\cap g^{-n+1}P_{\alpha_{n-1}}$. We denote this set $B_n(\rho)$. Fix a measure $\mu$ in $\mathcal{M}(S^*M,g^t)$. The Shannon-McMillan-Breiman Theorem states~\cite{Pa} that for $\mu$ almost $\rho$ in $S^*M$, the limit
$$\lim_{n+\infty}-\frac{1}{n}\log\mu(B_n(\rho))$$
 is well defined. We denote this limit $h_{KS}(\mu_{\rho},g,\mathcal{P})$. It defines an element in $L^1(\mu)$ that is $g^t$-invariant and that measures the exponential decrease of the $\mu$-volume of the ``$n$-balls'' $B_n(\rho)$. The Kolmogorov-Sinai entropy is then defined as
 $$h_{KS}(\mu,g)=\sup\left\{\int_{S^*M}h_{KS}(\mu_{\rho},g,\mathcal{P})d\mu(\rho):\mathcal{P}\ \text{is a finite partition of}\ S^*M\right\}.$$
Recall from the introduction that this quantity is affine for the ergodic decomposition and that it is bounded by the Ruelle-Margulis upper bound. Moreover, Abramov Theorem tells us that $h_{KS}(\mu,g^t)=|t|h_{KS}(\mu,g)$ for every $t$ in $\mathbb{R}$. Finally, underline that if there exists $C>0$ and $H_0\geq 0$ such that for every $n$ in $\mathbb{N}$ and for every $|\alpha|=n$, $\mu(P_{\alpha_0}\cap g^{-1}P_{\alpha_1}\ldots\cap g^{-n+1}P_{\alpha_{n-1}})\leq C e^{-nH_0}$, then $h_{KS}(\mu,g)\geq H_0.$

\subsection{Topological pressure} In corollary~\ref{c:loggap}, we made an assumption on the topological pressure of an invariant subset $K$ of $S^*M$. This quantity can be defined as the Legendre transform of the Kolmogorov-Sinai entropy~\cite{Pe, Wa}, i.e.
$$\forall f\in\mathcal{C}^0(S^*M),\ P_{top}\left(K,g^t,f\right):=\sup_{\mu\in\mathcal{M}(S^*M,g^t)}\left\{h_{KS}(\mu,g)+\int_Kfd\mu:\mu(K)=1\right\}.$$
This definition of topological pressure is known as the variational principle and we used this definition to derive corollary~\ref{c:loggap}. There are many other (equivalent) definitions of pressure based on thermodynamical formalism~\cite{Pe, Wa}. We just mention one of them here in order to clarify the statements of section~\ref{s:symboliccoding}. 

Given $\epsilon>0$ and $T\geq 0$, a subset $F$ of $K$ is said to be $(\epsilon,T)$-separated if for any $\rho\neq\rho'$ in $F$, there exists $0\leq t\leq T$ such that the distance $d(g^t\rho,g^t\rho')$ is $>\epsilon$. Fix now $f$ an element\footnote{In the following section, we will take $f=\frac{1}{2}\log J^u$.} in $\mathcal{C}^0(S^*M)$. Define
\begin{equation}\label{e:Tpressure}P_T(K,g^t, f,\epsilon)=\sup\left\{\sum_{\rho\in F}\exp\left(\int_0^Tf\circ g^t(\rho)dt\right)\right\},\end{equation}
where the supremum is taken over all $(\epsilon,T)$-separated subset of $K$. An equivalent definition of the topological pressure is then
$$P_{top}\left(K,g^t,f\right)=\lim_{\epsilon\rightarrow0}\limsup_{T\rightarrow+\infty}\frac{1}{T}\log P_T(K,g^t, f,\epsilon).$$

\section{Symbolic coding of the quantum dynamic}
\label{s:symboliccoding}

We fix $C$ a positive constant (that will be chosen small enough at the end of our proof). Let $(\psi_{\hbar})_{0<\hbar\leq \hbar_0}$ be a sequence of normalized vector in $L^2(M)$ such that
$$\mathcal{P}(\hbar,z)\psi_{\hbar}=z(\hbar)\psi_{\hbar},$$
where $z(\hbar)$ belongs to
\begin{equation}
 \label{e:spectralpara}
\left[\frac{1}{2}-\hbar, \frac{1}{2}+\hbar\right] +\imath\left[ -\frac{C\hbar}{|\log\hbar|},+\infty\right].
\end{equation}
Up to an extraction, we know that the distribution $\mu_{\psi_{\hbar}}$ defined by~(\ref{e:distrib}) converges weakly to a probability measure $\mu$. In the following, we 
will use the notation $\hbar\rightarrow0$ for the extraction in order to avoid heavy notations and in order to fit semiclassical notations.\\

A common point of~\cite{An08, AN07} is the introduction of a symbolic coding of the quantum dynamic to study localization properties of semiclassical measures. In these references, the proof relies on a careful study of the ``thermodynamical properties'' of the symbolic quantum system and on its link with the thermodynamical properties of semiclassical measures. We use also a symbolic presentation of the quantum system (similar to the one in~\cite{An08}) that we will describe in this section.

The main result of this section is proposition~\ref{p:crucialprop} and it will allow us to derive the proof of theorem~\ref{t:maintheo1} in section~\ref{s:proofs}. It shows, in a certain sense, that the sequence of eigenmodes defined above cannot concentrate on a subset of small topological pressure (at least for $C$ small enough).

\subsection{Energy cutoffs}\label{ss:cutoff}

As the sequence $(\psi_{\hbar})_{\hbar}$ concentrates on $S^*M$ in the semiclassical limit, we can introduce cutoff functions that will allow us to work with observables 
compactly supported in a small neighborhood of $S^*M$. First, we define $\tilde{\chi}$ a smooth function on $\mathbb{R}$ which is non negative, which is equal to $1$ for 
$|t|\leq 1/2$ and which is equal to $0$ for $|t|\geq 1$. Then, we fix $\delta>0$ and $k$ a positive integer. For every $0\leq j\leq k-1$, we define 
$$\forall\rho=(x,\xi)\in T^*M,\ \chi_{-j}(x,\xi):=\tilde{\chi}\left(4^j\delta^{-1}(\|\xi\|^2_x-1)\right).$$ 
Each of the function $\chi_{-j}$ is a smooth function on $S^*M$ which is compactly supported in $\{1-\delta/4^j\leq\|\xi\|^2_x\leq 1+\delta/4^j\}$. Moreover, by definition, 
the support of $1-\chi_{-j}$ and $\chi_{-j-1}$ are disjoint. Finally, for an eigenfunction $\psi_{\hbar}$, one has
$$\|\Op_{\hbar}(\chi_{-j})\psi_{\hbar}-\psi_{\hbar}\|=\mathcal{O}(\hbar^{\infty}),$$
thanks to the equality $\text{Id}-\Op_{\hbar}(\chi_{-j})=A_j(\mathcal{P}(\hbar,z)-z(\hbar))+R_j$ where $R_j$ is a smoothing operator and 
$A_j$ is a pseudodifferential operator of order $0$.

We underline that we will only work with a finite number of cutoffs (the integer $k$ we will take will only depend on $\delta$ and on a positive number $P_0$ 
with the notations of the following paragraphs).

\subsection{Smooth discretization of $M$}\label{s:discretization}

Let $M=M_1\sqcup\ldots M_{K}$ be a finite measurabe partition of $M$ of diameter bounded\footnote{We will fix $\epsilon$ small enough in a way that depends 
only on $M$ and on $P_0$ (see paragraph~\ref{ss:parameters}).} by $\frac{\epsilon}{2}$ and such that the measure $\mu$ does not charge the boundary of the partition. By lifting it on $T^*M$, it can be considered as a partition of $T^*M$. In~\cite{An08}, Anantharaman explained how to regularize such a partition in a smart way. Without getting into the details of~\cite{An08} (see paragraph $2.1$ and appendix $A.2$ of this reference), we recall that she constructed a family of smooth functions $(P_1^{\hbar},\ldots P_K^{\hbar})$ on $M$ (that depends on $\hbar$) satisfying in particular the following properties
\begin{itemize} 
 \item for every $i$, $P_i^{\hbar}\geq 0$;
 \item $\forall x\in M,\ \sum_{j=1}^KP_j^{\hbar}(x)=1$;
 \item $\overline{M_i}\subset\text{supp} P_i^{\hbar}\subset B(\overline{M_i},\frac{\epsilon}{4}),$
where $B(\overline{M_i},\frac{\epsilon}{4})$ is an $\epsilon/4$-neighborhood of $M_i$;
 \item $P_i^{\hbar}\rightarrow 1$ uniformly in every compact subset inside the interior of $M_i$, as $\hbar$ tends to $0$;
 \item $P_i^{\hbar}\rightarrow 0$ uniformly in every compact subset ouside $M_i$, as $\hbar$ tends to $0$;
 \item the growth of the derivatives is controlled by powers of $\hbar^{-\overline{\nu}}$ (with $\overline{\nu}<1/2$) and so the functions are amenable to $\hbar$-pseudodifferential calculus~\cite{DS, EZ} (see also appendix~\ref{a:pdo} for a brief reminder);
 \item There exists a $0<b<1/2$ such that
$$\forall i\neq j,\ \|P_i^{\hbar}P_j^{\hbar}\psi_{\hbar}\|=\mathcal{O}(\hbar^{\frac{b}{2}})\ \text{and}\ 
\forall i\neq j,\ \|(P_i^{\hbar})^2\psi_{\hbar}-P_i^{\hbar}\psi_{\hbar}\|=\mathcal{O}(\hbar^{\frac{b}{2}}).$$
\end{itemize}

The parameters $\overline{\nu}$ and $b$ are fixed in the following of the article.

For each of this smooth function $P_i^{\hbar}$, we can define a multiplication operator on $L^2(M)$
$$\forall u\in L^2(M),\ \pi_iu=P_i^{\hbar}\times u,$$
which is a bounded operator on $L^2(M)$ (of norm less than $1$). One can underline that
$$\sum_{i=1}^K\pi_i=\text{Id}_{L^2(M)}.$$
We introduce the following operator:
$$\forall\alpha\in \{1,\ldots, K\}^n,\ \Pi_{\alpha}:=\pi_{\alpha_{n-1}}(n-1)\ldots\pi_{\alpha_1}(1)\pi_{\alpha_0},$$
where $A(t):=\U_{\hbar}^{-t}A\U_{\hbar}^{t}.$ We underline that
$$\sum_{|\alpha|=n}\Pi_{\alpha}=\text{Id}_{L^2(M)}.$$

\subsection{Symbolic coding of the quantum dynamic}
\label{ss:partition}

Thanks to our smooth discretization of the manifold, we are now able to introduce a symbolic coding of the quantum dynamic  induced by $\U_{\hbar}^t$ and state our main result in terms of this 
symbolic dynamic.

\subsubsection{Quantum functionals on cylinders}
First, we define 
$$\Sigma:=\{1,\ldots, K\}^{\mathbb{N}}$$ 
and denote a cylinder $[\alpha_0,\ldots,\alpha_{n-1}]:=\{\forall 0\leq i\leq n-1, x_i=\alpha_i\}$. We will use $\beta.\alpha$ for the concatenation of two finite words 
$\beta:=(\beta_q,\ldots,\beta_{q+q'})$ and $\alpha:=(\alpha_p,\ldots,\alpha_{p+p'})$.

 We define the shift on $\Sigma$ as 
$\sigma((x_n)_{n\in\mathbb{N}})=(x_{n+1})_{n\in\mathbb{N}}$ and a \emph{quantum functional} on the cylinders of $\Sigma$
$$\mu_{\hbar}^{\Sigma}([\alpha_0,\ldots,\alpha_{n-1}])=\langle\Pi_{\alpha}\psi_{\hbar},\psi_{\hbar}\rangle_{L^2(M)}.$$

This object \textbf{is not a probability measure}. However, it satisfies the following nice properties 
\begin{prop}\label{p:probability} One has:
\begin{enumerate}
\item For every $n$ in $\mathbb{N}$, for every cylinder $[\alpha_0,\ldots,\alpha_{n-1}]$,
$$\sum_{\alpha_n}\mu_{\hbar}^{\Sigma}([\alpha_0,\ldots,\alpha_{n-1},\alpha_n])=\mu_{\hbar}^{\Sigma}([\alpha_0,\ldots,\alpha_{n-1}]);$$
\item For every $n$ in $\mathbb{N}$, for every cylinder $[\alpha_0,\ldots,\alpha_{n-1}]$ and for every $k$,
$$\mu_{\hbar}^{\Sigma}(\sigma^{-k}[\alpha_0,\ldots,\alpha_{n-1}])=\mu_{\hbar}^{\Sigma}([\alpha_0,\ldots,\alpha_{n-1}])+o_{n,k}(1);$$
\item For every $n$, 
$$\sum_{|\alpha|=n}\mu_{\hbar}^{\Sigma}([\alpha_0,\ldots,\alpha_{n-1}])=1.$$
\end{enumerate}
\end{prop}

In point $2$ of the proposition, we used the notation
$$\mu_{\hbar}^{\Sigma}(\sigma^{-k}[\alpha_0,\ldots,\alpha_{n-1}])=\sum_{\alpha_{-k},\ldots,\alpha_{-1}}\mu_{\hbar}^{\Sigma}([\alpha_{-k},\ldots,\alpha_{-1},\alpha_0,\ldots,\alpha_{n-1}]).$$

The ``quantum functional'' $\mu_{\hbar}^{\Sigma}$ looks very much like a $\sigma$-invariant probability measure. The two mains problems are that it is not positive a priori and that it is not exactly invariant.

Concerning the positivity, one can use Egorov property~(\ref{e:generalegorov}) with $q_1=\overline{\sqrt{2z(\hbar)}}V$ and $q_2=-\sqrt{2z(\hbar)}V$ and following the proof 
of proposition $1.3.2$ in~\cite{An08}, one finds that the following holds in the semiclassical limit $\hbar\rightarrow0$:
\begin{equation}\label{e:limitcyl}
\mu_{\hbar}^{\Sigma}([\alpha_0,\ldots,\alpha_{n-1}])\rightarrow\mu\left(g^{-n+1}M_{\alpha_{n-1}}\cap\ldots M_{\alpha_{0}}\right).
\end{equation}
So, in the semiclassical limit, $\mu_{\hbar}^{\Sigma}([\alpha_0,\ldots,\alpha_{n-1}])$ defines a nonnegative quantity.

Let us now explain how one can prove point $2$ of the proof. Using the fact that the semiclassical measure $\mu$ is $g^1$ invariant (as $\text{supp}(\mu)\subset \mathcal{N}_V$), one verifies that the 
limit of the ``quantum functional'' defines $\sigma$-invariant probability measure $\mu^{\Sigma}$ as follows:
$$\forall[\alpha_0,\ldots,\alpha_{n-1}],\ \mu^{\Sigma}([\alpha_0,\ldots,\alpha_{n-1}]):=\mu\left(M_{\alpha_0}\cap\ldots g^{-(n-1)}M_{\alpha_{n-1}}\right).$$
Hence, the functional becomes $\sigma$-invariant in the semiclassical limit. In particular, it implies point $2$ of the proposition, i.e. for fixed $n$ and $k$,
$$\mu_{\hbar}^{\Sigma}(\sigma^{-k}[\alpha_0,\ldots,\alpha_{n-1}])=\mu_{\hbar}^{\Sigma}([\alpha_0,\ldots,\alpha_{n-1}])+o_{n,k}(1).$$

As a conclusion, $\mu_{\hbar}^{\Sigma}$ does not define a $\sigma$-invariant probability measure for a fixed $\hbar$. However, in the semiclassical limit, it really defines a $\sigma$-invariant probability measure.

\begin{rema}

As mentionned in paragraph~\ref{pa:words}, the situation is slightly more complicated than the selfadjoint case treated in~\cite{An08} where the quantum functional was invariant under $\sigma$ (or at least invariant modulo a factor of order $k|\log\hbar|^{-1}$ in the case of quasimodes). Here, for a fixed $n$, there is a priori no reason to obtain a remainder $o_{k}(1)$ with a really explicit dependence in $k$. Yet, we will be able to prove a subinvariance property (proposition~\ref{p:subinvariance}) that will be sufficient for our proof.

For that purpose, we underline that the quantum functional $\mu_{\hbar}^{\Sigma}$ satisfies the following property that will be sufficient for our proof 
\begin{equation}\label{e:invariancepty}\mu_{\hbar}^{\Sigma}([\alpha_0,\ldots,\alpha_{n-1}])=e^{-\frac{2k\text{Im}(z(\hbar))}{\hbar}}\sum_{|\beta|=k}\left\langle(\U_{\hbar}^k)^*\U_{\hbar}^k\Pi_{\beta.\alpha}\psi_{\hbar},\psi_{\hbar}\right\rangle.
\end{equation}

\end{rema}

Define now the discrete analogue of $J^u$
$$J^{u}(\alpha_0,\alpha_1):=\sup\left(\left\{J^u(\rho):\rho\in M_{\alpha_0}\cap g^{-1}M_{\alpha_1},\|\rho\|\in[1-\delta,1+\delta]\right\}\cup\{\Lambda\}\right),$$
where $0<\Lambda\ll 1$. We also define $e^{\lambda_0}$ an upper bound on all the $J^{u}(\alpha_0,\alpha_1)$ that can be choose uniform for $\delta<1/2$ and for any choice of partition. For a given sequence $\alpha:=(\alpha_0,\ldots,\alpha_{n-1})$, we also define
$$J^{u}_n(\alpha):=J^{u}(\alpha_0,\alpha_1)\ldots J^u(\alpha_{n-2},\alpha_{n-1}).$$

Finally, we also introduce $\Sigma_n$ the set of $n$-cylinders in $\Sigma$, i.e.
$$\Sigma_n:=\{[\alpha_0,\ldots,\alpha_{n-1}]:\alpha_0,\ldots,\alpha_{n-1}\in\{1,\ldots,K\}\}.$$

\subsubsection{Main proposition}

The proof of the main theorem relies on the following proposition:

\begin{prop}\label{p:crucialprop}

Let $P_0$ be a positive number. There exist 
$$n_0'(P_0)\in\mathbb{N}\ \text{and}\ \epsilon_0(P_0)>0,$$
such that for every fixed choice of partition of diameter $\leq\epsilon_0(P_0)$, there exists
$$C(P_0)>0\ \text{and}\ 0\leq c(P_0)<1$$
such that if
\begin{itemize}
\item for $\hbar$ small enough, 
$$z(\hbar)\in \left[\frac{1}{2}-\hbar, \frac{1}{2}+\hbar\right] +\imath\left[ -C(P_0)\frac{\hbar}{|\log\hbar|},+\infty\right];$$
\item $n_0\geq n_0'(P_0)$;
\item $W_{n_0}$ is a subset of $\Sigma_{n_0}$ satisfying
\begin{equation}\label{e:hyppress}
\sum_{[\alpha]\in W_{n_0}}J^u_{n_0}(\alpha)^{\frac{1}{2}}\leq  e^{-\frac{n_0P_0}{2}},
\end{equation}
\end{itemize}
then, one has
$$\sum_{[\alpha]\in W_{n_0}}\mu^{\Sigma}([\alpha])\leq c(P_0).$$

\end{prop}

As was already mentioned, this proposition means that eigenmodes cannot concentrate entirely on cylinders satisfying a condition of ``small topological pressure''~(\ref{e:hyppress}). The sum appearing in~(\ref{e:hyppress}) is a discrete analogue of the quantities appearing in~\eqref{e:Tpressure}. Introducing these thermodynamical quantities generalizes slightly the strategy of~\cite{An08} where Anantharaman considered discrete versions of topological entropies (and not of topological pressures). 

We will explain in section~\ref{s:proofs} how one can deduce our main result from this proposition.

\subsection{Proof of proposition~\ref{p:crucialprop}}

We will admit several lemmas and show how they allow to derive proposition~\ref{p:crucialprop}. The proof of these intermediary lemmas will be given in section~\ref{s:intermediary}.

\begin{rema} 

Our proof requires the introduction of several small (or large) parameters. In order to avoid any confusion, we will summarize the links between the different parameters in paragraph~\ref{ss:parameters}.

\end{rema}

\subsubsection{Different scale of times}

In order to prove our result, we need to introduce various scale of times. The first one will be a fixed time $n_0\in\mathbb{N}$ that will play its main part at the \emph{classical level}. We fix $n_0$ a positive integer and a family $W_{n_0}$ of cylinders of length $n_0$ satisfying  

\begin{equation}\label{e:pressbound}
\sum_{[\alpha]\in W_{n_0}}J^u_{n_0}(\alpha)\leq e^{-\frac{P_0n_0}{2}}.
\end{equation}
The $n_0$ will be fixed large enough to apply lemma~\ref{l:counting}.

Fix now $\kappa>0$. The second important time is the so-called \emph{Ehrenfest time}
\begin{equation}\label{e:Ehrenfest}
n(\hbar):=\left[\kappa|\log\hbar|\right]. 
\end{equation}
This time is the one for which the 
semiclassical approximation will be valid for observables supported in a \emph{small macroscopic neighborhood} of $S^*M$ and for the observables $P_i^{\hbar}$. In particular, 
all the arguments of paragraphs~\ref{ss:strategy} and~\ref{p:conclusion} (and also of the appendix) will be valid for $0\leq p\leq n(\hbar)$. We underline that we will fix $\kappa$ small enough in a way that depends on 
$P_0$, on the partition $\mathcal{M}$, on the energy layer we work on and on the parameters $\overline{\nu}$ and $b$ used for the smoothing of the partition. 


Finally, we introduce $k\geq 2$ a large positive integer and 
a time that will be useful at the quantum level
\begin{equation}\label{e:quanttime}
N(\hbar)=kn(\hbar).
\end{equation}
In the following, $k$ and $\kappa$ will have to be chosen in a way that their product is bounded from below by a positive constant that will depend only on $P_0$ and on the dimension of $M$ (see paragraph~\ref{ss:intermed}).

In the following, we will omit the dependence of $N(\hbar)=N$ and $n(\hbar)=n$ in $\hbar$ to avoid heavy notations.

\subsubsection{Intermediary lemmas}\label{ss:intermed}

We start our proof by providing two intermediary lemmas that we will prove in section~\ref{s:intermediary}.\\

First, we fix $\tau$ in $[1/2,1]$ and we introduce a family of cylinders of length $p\geq n_0$ that spend a lot of time near $W_{n_0}$:
$$\Sigma_p(W_{n_0},\tau):=\left\{[\alpha]:=[\alpha_0,\ldots,\alpha_{p-1}]:\frac{\sharp\left\{j\in[0,p-n_0]:[\alpha_j,\ldots,\alpha_{j+n_0-1}]\in W_{n_0}\right\}}{p-n_0+1}\geq 
\tau\right\}.$$

In this paragraph, we will give lower and upper bounds on the following ``thermodynamical quantity'' associated to the family $\Sigma_p(W_{n_0},\tau)$:
$$\sum_{[\alpha]\in \Sigma_p(W_{n_0},\tau)}J^u_p(\alpha)^{\frac{1}{2}}=\sum_{[\alpha]\in \Sigma_p(W_{n_0},\tau)}\exp\left(\frac{1}{2}\sum_{j=0}^{p-1}\log J^u(\alpha_j,\alpha_{j+1})\right).$$
Roughly speaking, we will verify that if ``the eigenmodes put too much weight'' on the cylinders $\Sigma_n(W_{n_0},\tau)$, then this thermodynamical quantity must grow faster than it is authorized by the thermodynamical assumption~\eqref{e:pressbound} on $W_{n_0}$. In particular, it will show that eigenmodes cannot concentrate entirely on $\Sigma_n(W_{n_0},\tau)$.

The first lemma provides a general upper bound on this ``thermodynamical quantity'' that relies only on the thermodynamical assumption~\eqref{e:pressbound} on $W_{n_0}$:

\begin{lemm}\label{l:counting} 
There exists $n_0'$ and $p_0$ depending only on $P_0$ such that for every $p\geq p_0$, for every $n_0\geq n_0'$, for every $W_{n_0}$ satisfying~(\ref{e:pressbound}) and for every $\tau\in[1/2,1]$,
$$\sum_{[\alpha]\in\Sigma_p(W_{n_0},\tau)}J^u_p(\alpha)^{\frac{1}{2}}\leq e^{-\frac{pP_0}{8}+p(1-\tau)(\lambda_0+\log K)} (e^{\lambda_0}K)^{n_0}e^{\frac{n_0P_0}{2}}.$$
 
\end{lemm}

We recall that $K$ is the cardinal of the partition of $M$ and $e^{\lambda_0}$ is an upper bound on all the $J^u(\alpha_0,\alpha_1)$.

This lemma relies only on the \emph{classical properties} of the system. In particular, if we apply this lemma for $p=n$, $n_0\geq n_0'$ and for $\tau\geq \frac{1}{2}$, we find that, for $\hbar$ small 
enough, one has
\begin{equation}\label{e:upboundpress}
\sum_{[\alpha]\in\Sigma_n(W_{n_0},\tau)}J^u_n(\alpha)^{\frac{1}{2}}\leq e^{\left(-\frac{P_0}{8}+(1-\tau)(\lambda_0+\log K)\right)\kappa|\log\hbar|}
e^{\frac{P_0}{8}}(e^{\lambda_0+\frac{P_0}{2}}K)^{n_0}.
\end{equation}
In particular, this last equality combined to assumption~(\ref{e:spectralpara}) implies that
\begin{equation}\label{e:upboundpress2}\sum_{[\alpha]\in\Sigma_n(W_{n_0},\tau)}J^u_n(\alpha)^{\frac{1}{2}}e^{-\frac{(n-1)\text{Im}(z(\hbar))}{\hbar}}=\mathcal{O}\left(\hbar^{\kappa(-(1-\tau)(\lambda_0+\log K)
+\frac{P_0}{8})}\right),\ \text{as}\ \hbar\rightarrow0.\end{equation}

As explained above, the second lemma gives a lower bound on $\sum_{[\alpha]\in\Sigma_n(W_{n_0},\tau)}J^u_n(\alpha)^{\frac{1}{2}}$ under an assumption on the concentration of the 
eigenmodes $(\psi_{\hbar})_{\hbar}$. Precisely, we show that if the eigenfunction charges the cylinders in $\Sigma_n(W_{n_0},\tau)$, then the previous sum is bounded from below by a precise 
power of $\hbar$:

\begin{lemm}\label{l:lowbound} Let $k$, $\kappa$ and $C$ be as above\footnote{The constant $C>0$ is the one given by the spectral window~(\ref{e:spectralpara}).}. Let $\theta$ be an element in $[0,1]$. If $W_{n_0}$ is a family of $n_0$-cylinders satisfying~(\ref{e:pressbound}) and if
$$\left\|\sum_{[\alpha]\in\Sigma_n(W_{n_0},\tau)^c}\pi_{\alpha_{n-1}}\U_{\hbar}\ldots\pi_{\alpha_1}\U_{\hbar}\pi_{\alpha_0}\psi_{\hbar}\right\|_{L^2(M)}\leq e^{\frac{(n-1)\text{Im}(z(\hbar))}{\hbar}}
\theta \frac{e^{-Ck\kappa}}{k},$$
where $\Sigma_n(W_{n_0},\tau)^c$ means the complementary of $\Sigma_n(W_{n_0},\tau)$ in $\Sigma_n$; then, one has

 $$\left(\sum_{[\alpha]\in \Sigma_n(W_{n_0},\tau)}J^u_n(\alpha)^{\frac{1}{2}}e^{-\frac{(n-1)\text{Im}(z(\hbar)}{\hbar}}\right)^k\hspace{7cm}$$
\begin{equation}\label{e:lowboundpress}\geq e^{-k\lambda_0-\frac{(k-1)\text{Im}(z(\hbar)}{\hbar}} 
\left(1-(1+\mathcal{O}(\hbar^{\nu_0'}))^k\theta
+\mathcal{O}(\hbar^{\infty})\right)\hbar^{d/2}e^{-C_0k\kappa\epsilon|\log\hbar|},
\end{equation}
where $\nu_0'>0$ and $\mathcal{O}(\hbar^{\nu_0'})$ depend on $\kappa$ and on the smooth partition and $C_0$ depends only on $M$ and on $\delta>0$ (the size of the energy layer).

\end{lemm}

We recall that $\epsilon$ is an upper bound on the diameter of the partition

We will prove these two intermediary lemmas in section~\ref{s:intermediary}. We explain in the next paragraphs how they allow to derive proposition~\ref{p:crucialprop}.

\subsubsection{Using the intermediary lemmas}

Lemma~\ref{l:lowbound} shows that under an assumption on the concentration of the eigenmodes, one has a lower bound on 
$$\sum_{[\alpha]\in \Sigma_n(W_{n_0},\tau)}J^u_n(\alpha)^{\frac{1}{2}}e^{-\frac{(n-1)\text{Im}(z(\hbar)}{\hbar}}.$$
Precisely, it tells us that this lower bound is of order $\hbar^{\frac{d}{2k}+C_0\kappa\epsilon}.$ We can now compare this lower bound to the upper bound~(\ref{e:upboundpress2}). If we are able to take the different parameters in a way that
\begin{equation}\label{e:linkparameter}\frac{d}{2k}+C_0\kappa\epsilon< \kappa\left(-(1-\tau)(\lambda_0+\log K)
+\frac{P_0}{8}\right),\end{equation}
then we will have that
$$\left\|\sum_{[\alpha]\in\Sigma_n(W_{n_0},\tau)^c}\pi_{\alpha_{n-1}}\U_{\hbar}\ldots\pi_{\alpha_1}\U_{\hbar}\pi_{\alpha_0}\psi_{\hbar}\right\|_{L^2(M)}\geq e^{\frac{(n-1)\text{Im}(z(\hbar))}{\hbar}}
\theta \frac{e^{-Ck\kappa}}{k}.$$

In order to obtain relation~(\ref{e:linkparameter}), we first choose $\epsilon$ small enough (depending only on $P_0$). Then, we choose $\tau_0<1$ (depending 
also on the partition) close enough to $1$ to have
$$-(1-\tau_0)(\lambda_0+\log K)
+\frac{P_0}{8}-C_0\epsilon\geq \frac{P_0}{16}$$
We underline that this inequality remains true for any $\tau\geq \tau_0$. Finally, we can take any $k$ and $\kappa$ satisfying $k\kappa> \frac{8d}{P_0}$. So the more $\kappa$ will be small, the more we will have to take $k$ large.

For this choice of small parameters, an eigenmode cannot put all his weight on the cylinders in $\Sigma_n(W_{n_0},\tau)$, i.e. on a subset of small topological pressure with respect to $\frac{1}{2}\log J^u$.

\subsubsection{Using the semiclassical approximation}\label{ss:strategy}

In the previous paragraph, we saw that for a partition of small enough diameter and for $\tau$ close enough to $1$, one has, for $\hbar$ small enough,
\begin{equation}\label{e:scbound}\left\|\sum_{[\alpha]\in\Sigma_n(W_{n_0},\tau)^c}\pi_{\alpha_{n-1}}\ldots\pi_{\alpha_1}(2-n)\pi_{\alpha_0}(1-n)\psi_{\hbar}\right\|_{L^2(M)}^2\geq
\theta^2 \frac{e^{-2Ck\kappa}}{k^2}.\end{equation}
We would like now to relate this lower bound to a lower bound on
$$\sum_{[\alpha]\in\Sigma_n(W_{n_0},\tau)^c}\mu_{\hbar}^{\Sigma}([\alpha]).$$
This can be achieved using the fact that for $\kappa$ small enough (depending on the choice of the partition), the pseudodifferential operators we consider are amenable to semiclassical calculus. Moreover, thanks to our choice of partition, one knows that the family $(\pi_i)_i$ forms a family of almost orthogonal projectors when it acts on the eigenmodes $\psi_{\hbar}$ (see paragraph~\ref{s:discretization}). At this point, we face the same difficulty as in paragraph $2.4$ of~\cite{An08} and we will briefly recall in paragraph~\ref{p:semiclassicalapprox} how one can prove that, for $\kappa$ small enough,
$$\left\|\sum_{[\alpha]\in\Sigma_n(W_{n_0},\tau)^c}\pi_{\alpha_{n-1}}\ldots\pi_{\alpha_1}(2-n)\pi_{\alpha_0}(1-n)\psi_{\hbar}\right\|_{L^2(M)}^2\hspace{3cm}$$
$$=\sum_{[\alpha]\in\Sigma_n(W_{n_0},\tau)^c}\left\langle\pi_{\alpha_{n-1}}\ldots\pi_{\alpha_1}(2-n)\pi_{\alpha_0}(1-n)\psi_{\hbar},\psi_{\hbar}\right\rangle
+\mathcal{O}(\hbar^{\nu_0'}),$$
where $\nu_0'$ is a positive constant that depends on the partition $\mathcal{M}$, on the energy layer we work on and on the parameters $\overline{\nu}$ and $b$ used for the smoothing of the partition (see paragraphs~\ref{ss:cutoff} and~\ref{s:discretization}).

\begin{rema}
In the following, we will also use the exponent $\nu_0'$ for the other remainders due to the semiclassical approximation, meaning that we will always keep the worst remainder term.
\end{rema}

Using the fact that $\psi_{\hbar}$ is an eigenmode, we also find
$$\left\|\sum_{[\alpha]\in\Sigma_n(W_{n_0},\tau)^c}\pi_{\alpha_{n-1}}\ldots\pi_{\alpha_1}(2-n)\pi_{\alpha_0}(1-n)\psi_{\hbar}\right\|_{L^2(M)}^2\hspace{3cm}$$
$$=\sum_{[\alpha]\in\Sigma_n(W_{n_0},\tau)^c}e^{-\frac{2(n-1)\text{Im}(z(\hbar))}{\hbar}}\left\langle(\U_{\hbar}^{n-1})^*\U_{\hbar}^{n-1}\Pi_{\alpha}\psi_{\hbar},\psi_{\hbar}\right\rangle+\mathcal{O}(\hbar^{\nu_0'}).$$

We have to face here a problem which is due to the ``nonselfadjointness'' of our problem. As in the simple example of paragraph~\ref{pa:words}, we will use the fact that the quantization is almost positive (more precisely lemma~\ref{l:subinvariance}). Moreover, we can use the fact that there is at most $K^n$ terms in the sum. Combining these two properties to lower bound~(\ref{e:scbound}), one finally gets

$$\theta^2 \frac{e^{-2Ck\kappa}}{k^2}e^{2(n-1)\frac{\text{Im}\ z(\hbar)}{\hbar}}\leq 
\sum_{[\alpha]\in\Sigma_n(W_{n_0},\tau)^c}\mu_{\hbar}^{\Sigma}([\alpha])+\mathcal{O}(\hbar^{\nu_0'}).$$
Thanks to the property of partition of identity, one also has
\begin{equation}\label{e:mainestimate}
\sum_{[\alpha]\in\Sigma_n(W_{n_0},\tau)}\mu_{\hbar}^{\Sigma}([\alpha])\leq \left(1-\theta^2 e^{2(n-1)\frac{\text{Im}\ z(\hbar)}{\hbar}}\frac{e^{-2Ck\kappa}}{k^2}\right)+\mathcal{O}(\hbar^{\nu_0'}).
\end{equation}

\subsubsection{The conclusion: from time $n(\hbar)$ to time $n_0$}\label{p:conclusion}

In the upper bound~\eqref{e:mainestimate}, $n$ and $\mu_{\hbar}^{\Sigma}$ depend both on $\hbar$; hence, one cannot directly take the limit $\hbar\rightarrow0$ and derive 
proposition~\ref{p:crucialprop}. We will now show how we can derive an 
estimate on $\mu_{\hbar}^{\Sigma}(W_{n_0})$, where we use the notation
$$\forall W\subset\Sigma_p,\ \mu_{\hbar}^{\Sigma}(W):=\sum_{[\alpha]\in W}\mu_{\hbar}^{\Sigma}([\alpha]).$$

Using proposition~\ref{p:subinvariance}, we write
$$\mu_{\hbar}^{\Sigma}(W_{n_0})\leq\frac{e^{-2(n-1)\frac{\text{Im}\ z(\hbar)}{\hbar}}}{n-n_0}\sum_{k=0}^{n-n_0-1}\sum_{|\beta|=k}\sum_{[\alpha]\in W_{n_0}} 
\mu_{\hbar}^{\Sigma}([\beta.\alpha])+\mathcal{O}_{n_0}(\hbar^{\nu_0'}),$$
where $\nu_0'$ has the same properties as above. Then, one can use the compatibility relation of proposition~\ref{p:probability} to derive
\begin{equation}\label{e:birkaverage}\mu_{\hbar}^{\Sigma}(W_{n_0})\leq e^{-2(n-1)\frac{\text{Im}\ z(\hbar)}{\hbar}}\mu_{\hbar}^{\Sigma}\left(\frac{1}{n-n_0}\sum_{k=0}^{n-n_0-1}\mathbf{1}_{\sigma^{-k}W_{n_0}}\right)+\mathcal{O}_{n_0}(\hbar^{\nu_0'}).\end{equation}

We can now proceed as in~\cite{An08}, i.e. combine the following observations:
\begin{itemize}
\item on $\Sigma_n(W_{n_0},\tau)$, one has
$$\frac{1}{n-n_0}\sum_{k=0}^{n-n_0-1}\mathbf{1}_{\sigma^{-k}W_{n_0}}\leq 1,$$
and, on $\Sigma_n(W_{n_0},\tau)^c$,
$$\frac{1}{n-n_0}\sum_{k=0}^{n-n_0-1}\mathbf{1}_{\sigma^{-k}W_{n_0}}\leq \tau;$$
\item the length of the cylinders involved in~\eqref{e:birkaverage} is of order $\kappa|\log\hbar|$, with $\kappa$ small enough to have the operators amenable to semiclassical calculus and the functional $\mu_{\hbar}^{\Sigma}$ is almost positive at least on these cylinders of length -- see paragraph~\ref{p:antiwick} and lemma~\ref{l:productPDO};
\item there are $K^n$ terms in the sums involved in the upper bound~\eqref{e:birkaverage}.
\end{itemize}
These properties allows to find, for $\kappa$ small enough,
$$\mu_{\hbar}^{\Sigma}(W_{n_0})\leq e^{-2(n-1)\frac{\text{Im}\ z(\hbar)}{\hbar}}\left(\tau\mu_{\hbar}^{\Sigma}\left(\Sigma_n(W_{n_0},\tau)^c\right)+\mu_{\hbar}^{\Sigma}\left(\Sigma_n(W_{n_0},\tau)\right)\right)+\mathcal{O}(\hbar^{\nu_0'}),$$
where $\nu_0'>0$. Thanks to the property of partition of identity, one can verify that
$$\mu_{\hbar}^{\Sigma}(W_{n_0})\leq e^{-2(n-1)\frac{\text{Im}\ z(\hbar)}{\hbar}}\left(\tau+(1-\tau)\mu_{\hbar}^{\Sigma}\left(\Sigma_n(W_{n_0},\tau)\right)\right)+\mathcal{O}(\hbar^{\nu_0'}).$$
At this point of the proof one can use our assumption on the quantum decay rate $\frac{2\text{Im}\ z(\hbar)}{\hbar}$. In fact, if we implement property~\eqref{e:spectralpara} in inequality~(\ref{e:mainestimate}) and if we let $\hbar$ tends to $0$, then we derive
$$\mu^{\Sigma}(W_{n_0})\leq e^{2C\kappa}\left(\tau+(1-\tau)\left(1- \frac{\theta^2e^{-2C(k+1)\kappa}}{k^2}\right)\right).$$
This inequality holds for any $\theta$ in $(0,1)$ and hence,
\begin{equation}\label{e:finalstep}\mu^{\Sigma}(W_{n_0})\leq e^{2C\kappa}\left(\tau+(1-\tau)\left(1- \frac{e^{-2C(k+1)\kappa}}{k^2}\right)\right).
\end{equation}
As all the other paramaters were fixed before and as this inequality holds for any $C>0$, one can now take $C$ small enough to have 
$$e^{2C\kappa}\left(\tau+(1-\tau)\left(1- \frac{e^{-2C(k+1)\kappa}}{k^2}\right)\right)<1.$$
It concludes the proof of proposition~\ref{p:crucialprop}.

\subsection{Comments on the choice of the different parameters}\label{ss:parameters} In order to avoid any confusion, we summarize here all the relations between the different parameters. 

First, as the eigenfunctions concentrate on $S^*M$, we have fixed a small neighborhood of $S^*M$ of size $0<\delta<1/2$ (``the energy layer'') and we fixed $\overline{\nu}<1/2$ and $0<b<1/2$ that we will use to define the regularized partition.

We also fixed a real positive number $P_0$.

Then, we have introduced a partition of small diameter depending only on $P_0$. Once a partition is fixed, we can introduce symbolic coding of the quantum dynamic. 
We also fix $\tau>1$ larger than some $\tau_0$ depending on $P_0$ and on the partition (see paragraph~\ref{ss:intermed}).

Once these parameters are fixed, we fix a parameter $\kappa>0$ small enough depending on $P_0$, on $\delta$, on the partition $\mathcal{M}$ and on the parameters $\overline{\nu}$ and $b$ used for the smoothing of the partition. It is small enough to make the arguments of paragraphs~\ref{ss:strategy},~\ref{p:conclusion} and also of the appendix work. Then, we fix $k$ large enough to have $k\kappa >\frac{8 d}{P_0}$.

Finally, once all these parameters are fixed, we fix some $C$ small enough such that the sum in the upper bound of~(\ref{e:finalstep}) is strictly less than $1$.

\section{Proof of Theorem~\ref{t:maintheo1}}

\label{s:proofs}

In the previous section, we gave a symbolic description of the quantum system and of its semiclassical limit. This symbolic coding is a standard procedure in ergodic theory and we 
will now show how one can relate the results obtained in the symbolic setting to the main Theorem of the introduction. Precisely, we will show how proposition~\ref{p:crucialprop} implies theorem~\ref{t:maintheo1} following an argument from~\cite{An08}.\\

Let $P_0$ be a positive number and let $\mathcal{M}:=(M_i)_{i=1}^K$ be a partition of $M$ with small diameter as in  proposition~\ref{p:crucialprop}. Let $\rho$ be an element in $S^*M$ and $n_0$ be positive integer. There exists an unique 
$[\alpha_{\rho}]=[\alpha_0,\ldots,\alpha_{n_0-1}]$ in $\Sigma_{n_0}$ such that $\rho$ belongs to $M_{\alpha_0}\cap g^{-1}M_{\alpha_1}\cap\ldots g^{-n_0+1} M_{\alpha_{n_0-1}}$. We denote this subset $M_{n_0}(\rho)$. One can remark that if we choose the diameter of the partition $\epsilon$ small enough\footnote{Underline that $(M_{\alpha_0}\cap g^{-1}M_{\alpha_1})_{\alpha_0,\alpha_1}$ 
defines then a partition of small diameter of $S^*M$.} and the size of the energy layer $\delta$ small enough, then for every $\rho\in S^*M$, one has  
$$\left|\frac{1}{2}\sum_{j=0}^{n_0-1}\log J^u\circ g^j(\rho)-\frac{1}{2}\log J^u_{n_0}\left(\alpha_{\rho}\right)\right|\leq \frac{n_0 P_0}{2}.$$

Now, we consider $\mu$ in $\mathcal{M}((\psi_{\hbar})_{0<\hbar\leq\hbar_0})$ as in the statement of theorem~\ref{t:maintheo1} where $C(P_0)$ is given by proposition~\ref{p:crucialprop}. We write the ergodic decomposition of $\mu$~\cite{EinLin06} 
$$\mu=\int_{S^*M}\mu_{\rho}d\mu(\rho),$$
where every $\mu_{\rho}$ is an ergodic probability measure. According to section~\ref{s:background}, one also knows that, for $\mu$ almost every $\rho\in S^*M$,
\begin{equation}\label{e:Lyapunov}
\frac{1}{n_0}\sum_{j=0}^{n_0-1}\log J^u\circ g^j(\rho)=\frac{1}{n_0}\log J^u_{n_0}(\rho)\longrightarrow \int_{S^*M}\log J^ud\mu_{\rho},\ \text{as}\ n_0\rightarrow+\infty.
\end{equation}

Introduce now
$$I_{P_0}:=\left\{\rho\in S^*M:h_{KS}(\mu_{\rho},g,\mathcal{M})<-\frac{1}{2}\int_{S^*M}\log J^ud\mu_{\rho}-2P_0\right\}.$$
According to the Shannon-McMillan-Breiman Theorem~\cite{Pa}, one knows that for $\mu$ almost every $\rho$ in $S^*M$,
\begin{equation}\label{e:Breiman}
h_{KS}(\mu_{\rho},g,\mathcal{M})=\lim_{n_0\rightarrow +\infty}-\frac{1}{n_0}\log\mu (M_{n_0}(\rho)).
\end{equation}
We will now prove by contradiction that 
$$\mu(I_{P_0})\leq c(P_0),$$ 
where $0\leq c(P_0)<1$ is the constant that appears in proposition~\ref{p:crucialprop}. Suppose $\mu(I_{P_0})> c(P_0)$.

One knows that, for every $\eta>0$, there exists $\mathcal{R}$ such that $\mu(\mathcal{R})\leq\eta$ and 
such that the previous limits~(\ref{e:Breiman}) and~(\ref{e:Lyapunov}) holds uniformly for $\rho\in I_{P_0}-\mathcal{R}$. Denote $I_{P_0}^{\eta}=I_{P_0}-\mathcal{R}$. 
According to the definition of $I_{P_0}^{\eta}$, one knows 
that there exists $n_0'(P_0,\eta)$ such that for every $n_0\geq n_0'(P_0,\eta)$ and for every $\rho\in I_{P_0}^{\eta}$, one has
$$-\log\mu (M_{n_0}(\rho))+\frac{1}{2}\sum_{j=0}^{n_0-1}\log J^u\circ g^j(\rho)\leq -P_0 n_0.$$

This implies that for every $\rho$ in $I_{P_0}^{\eta}$, one has
$$ J^u_{n_0}\left(\alpha_{\rho}\right)^{\frac{1}{2}}\leq e^{-\frac{n_0P_0}{2}}\mu (M_{n_0}(\rho)).$$
Consider now $F$ a finite subset of $I_{P_0}^{\eta}$ such that $\rho\neq\rho'$ in $F$ implies that $M_{n_0}(\rho)\neq M_{n_0}(\rho')$ and such that 
$$I_{P_0}^{\eta}\subset\bigsqcup_{\rho\in F}M_{n_0}(\rho).$$
One has that $\displaystyle\sum_{\rho\in F}  J^u_{n_0}\left(\alpha_{\rho}\right)^{\frac{1}{2}}\leq e^{-\frac{n_0P_0}{2}}.$ According to proposition~\ref{p:crucialprop}, one knows that
$$\mu(I_{P_0}^{\eta})\leq\mu\left(\bigcup_{\rho\in F}M_{n_0}(\rho)\right)\leq c(P_0).$$
This inequality holds for every $\eta>0$ and we obtain the contradiction for $\eta$ small enough. Finally, we have
$$\mu\left(\left\{\rho\in S^*M:h_{KS}(\mu_{\rho},g,\mathcal{M})\geq-\frac{1}{2}\int_{S^*M}\log J^ud\mu_{\rho}-2P_0\right\}\right)\geq
1-c(P_0)>0.$$
It concludes the proof of Theorem~\ref{t:maintheo1}.

\section{Proof of intermediary lemmas}
\label{s:intermediary}

In this section, we give the proof of intermediary results that were at the heart of our proof of proposition~\ref{p:crucialprop}.

\subsection{Proof of lemma~\ref{l:counting}}

The proof of this lemma relies only on the classical properties of our problem (and not on its quantum structure). As mentionned above, we fix $n_0$ a positive integer and a family $W_{n_0}$ of cylinders of length $n_0$ such that
$$\sum_{[\alpha]\in W_{n_0}}J^u_{n_0}(\alpha)^{\frac{1}{2}}\leq e^{-\frac{P_0n_0}{2}}.$$
Our only assumption on $n_0$ is that it is large enough so that for $p$ large enough, one has
$$\left(\begin{array}{c}p \\ \left[\frac{p}{n_0}\right]\end{array}\right)\leq e^{p\frac{P_0}{16}}.$$
We underline that $n_0$ is large in a way that depends only on $P_0$.\\

The proof follows a similar strategy as its analogue in~\cite{An08}, $\S 2.3$, except that we consider slightly different dynamical quantities related to dynamical pressures and not to entropies.

\subsubsection*{Decomposition of cylinders in $\Sigma_p(W_{n_0},\tau)$}

Let $[\alpha]=[\alpha_0,\ldots,\alpha_{n-1}]$ be an element in $\Sigma_p(W_{n_0},\tau)$. We will first show that $\alpha$ can be decomposed into the concatenation of well-chosen 
cylinders. In order to describe this decomposition, we introduce an increasing sequence of stopping times. First, one sets
$$t_0:=\inf\left\{0\leq j\leq p-n_0:[\alpha_j,\ldots,\alpha_{j+n_0-1}]\in W_{n_0}\right\}.$$ 
As $\tau\geq 1/2$, one knows from the definition of $\Sigma_p(W_{n_0},\tau)$ that $t_0$ is well defined. Then, as long as the induction 
is well defined, we define the following increasing sequence of integers:
$$t_1:=\inf\left\{t_0+n_0\leq j\leq p-n_0:[\alpha_j,\ldots,\alpha_{j+n_0-1}]\in W_{n_0}\right\},\ldots$$
$$t_{l+1}:=\inf\left\{t_l+n_0\leq j\leq p-n_0:[\alpha_j,\ldots,\alpha_{j+n_0-1}]\in W_{n_0}\right\}.$$ 
We associate to this sequence a sequence of intervals of length $n_0$
$$I_0=[t_0,t_0+n_0-1],\ldots,I_l=[t_l,t_l+n_0-1].$$

From the definition of our sequence, one knows that, for $0\leq j\leq t_{l}+n_0-1$ outside $\cup_{j'}I_{j'}$ (also for $j\leq p-n_0$), $[\alpha_j,\ldots,\alpha_{j+n_0-1}]$ does 
not belong to $W_{n_0}$. In particular, as ones knows that there are at most $(1-\tau)(p-n_0)$ such $j$. The stopping times are well defined at least for every $l\leq\frac{\tau p}{n_0}
-\tau$. Hence, one can define $l$ disjoint intervals as above with $\frac{p}{2n_0}-1\leq\frac{\tau p}{n_0}-\tau\leq l\leq\frac{p}{n_0}$ (remind that we chose $1/2\leq\tau\leq 1$).

In the end, we find that $[\alpha]$ in $\Sigma_p(W_{n_0},\tau)$ can be written as $[b_0;c_0;\ldots;b_{l-1};c_{l-1};b_{l}]$ 
where
\begin{itemize}
 \item every subcylinder $c_j$ belongs to $W_{n_0}$;
 \item every subcylinder $b_j$ (if not empty) contains letters $\alpha_k$ such that $[\alpha_k,\ldots,\alpha_{k+n_0-1}]$ (when it makes sense, i.e. $k\leq p-n_0$) does not 
belong to $W_{n_0}$.
\end{itemize}

From the definition of $\Sigma_p(W_{n_0},\tau)$, one knows that $\sum_{j=0}^{l}|b_j|\leq (1-\tau)(p-n_0)+n_0\leq(1-\tau)p-n_0$.

\subsubsection*{Upper bound on $\sum_{[\alpha]\in \Sigma_p(W_{n_0},\tau)}J^u_p(\alpha)^{\frac{1}{2}}$}

From the previous paragraph, a cylinder $[\alpha_0,\ldots,\alpha_{p-1}]$ in $\Sigma_p(W_{n_0},\tau)$ is determined by the following data:
\begin{enumerate}
 \item the subcylinders $(a_j)_{0\leq j\leq l-1}$ where $\frac{p}{2n_0}-1\leq l\leq p/n_0$;
 \item the subcylinders $(b_j)_{0\leq j\leq l}$ where $l\leq p/n_0$. 
\end{enumerate}

Regarding this decomposition, we will now give an upper bound on 
$$\sum_{[\alpha]\in \Sigma_p(W_{n_0},\tau)}J^u_p(\alpha)^{\frac{1}{2}}=\sum_{[\alpha]\in \Sigma_p(W_{n_0},\tau)}J^u(\alpha_0,\alpha_1)^{\frac{1}{2}}\ldots
J^u(\alpha_{n-2},\alpha_{n-1})^{\frac{1}{2}}.$$

For a fixed choice of positions for the subcylinders $(b_j)_{0\leq j\leq l'}$, the number of possibility for the values of $(b_j)_{0\leq j\leq l}$ is bounded by $K^{(1-\tau)p+n_0}$, where $K$ is the cardinal of the partition. In this case, 
we bound $J^u(\alpha_i,\alpha_{i+1})^{\frac{1}{2}}$ by $e^{\lambda_0}$.

Choosing a family of positions for the subcylinders $(a_j)$ is equivalent to choose a family of endpoints; so there are at most 
$\displaystyle\left(\begin{array}{c}p \\ \left[\frac{p}{n_0}\right]\end{array}\right)^2$ choices of family of position for the subcylinders $a_j$. For a given position of 
these subcylinders, the sum runs (for every $a_j$) over the cylinders in $W_{n_0}$, for which one has
$$\sum_{\alpha\in W_{n_0}}J^u_{n_0}(\alpha)^{\frac{1}{2}}\leq e^{-\frac{n_0P_0}{2}}.$$

Finally, we find that
$$\sum_{[\alpha]\in \Sigma_p(W_{n_0},\tau)}J^u_p(\alpha)^{\frac{1}{2}}\leq \left(\begin{array}{c}p \\ \left[\frac{p}{n_0}\right]\end{array}\right)^2 
(e^{\lambda_0}K)^{(1-\tau)p+n_0}e^{\frac{n_0P_0}{2}}e^{-\frac{pP_0}{4}} ,$$
which concludes the proof of the lemma thanks to our assumption on $n_0$.

\subsection{Proof of lemma~\ref{l:lowbound}}

In order to prove lemma~\ref{l:lowbound}, we introduce a family of cylinders of length $N=kn$ related to $\Sigma_n(W_{n_0},\tau)$, where $k$ is a fixed integer.

Precisely, we define 
$\Sigma_n(W_{n_0},\tau)^k$ as the family of cylinders of the form $[\gamma]:=[\gamma^0;\ldots;\gamma^{k-1}]$ where every $\gamma_j$ is an element of $\Sigma_n(W_{n_0},\tau)$. We use the notation $(\Sigma_n(W_{n_0},\tau)^k)^c$ for its complementary in $\Sigma_N$. We will prove lemma~\ref{l:lowbound} using the following lemma:

\begin{lemm}[submultiplicativity property]\label{l:submultip} Fix $\theta$
in $[0,1]$. If
$$\left\|\sum_{[\alpha]\in\Sigma_n(W_{n_0},\tau)^c}\pi_{\alpha_{n-1}}\U_{\hbar}\ldots\pi_{\alpha_1}\U_{\hbar}\pi_{\alpha_0}\psi_{\hbar}\right\|_{L^2(M)}\leq \frac{e^{-Ck\kappa}}{k}\theta
e^{\frac{(n-1)\text{Im}(z(\hbar))}{\hbar}},$$
 where $\Sigma_n(W_{n_0},\tau)^c$ means the complementary of $\Sigma_n(W_{n_0},\tau)$ in $\Sigma_n$; then, one has
$$\left\|\sum_{[\gamma]\in (\Sigma_n(W_{n_0},\tau)^k)^c}\pi_{\gamma_{N-1}}\U_{\hbar}\ldots\pi_{\gamma_1}\U_{\hbar}\pi_{\gamma_0}\psi_{\hbar}\right\|\hspace{5cm}$$
$$\leq
\theta  \left(1+\mathcal{O}(\hbar^{\nu_0'})\right)^{k}
 e^{\frac{(N-1)\text{Im}(z(\hbar))}{\hbar}},$$
where $\nu_0'>0$ and $\mathcal{O}(\hbar^{\nu_0'})$ depend on $\kappa$ and on the smooth partition.

\end{lemm}

This lemma is a crucial step as it allows to connect relations on the scale of the Ehrenfest time $n(\hbar)$ to relations on the scale $kn(\hbar)$, where $k$ is an arbitrary integer. We postpone the proof of this lemma to the end of this paragraph and first show how we can derive lemma~\ref{l:lowbound} from it.

\subsubsection{Combining lemma~\ref{l:submultip} to hyperbolic dispersive estimates from~\cite{An08}}

In order to prove lemma~\ref{l:lowbound}, one can write that
$$\sum_{[\gamma]\in (\Sigma_n(W_{n_0},\tau)^k)^c}\langle\pi_{\gamma_{N-1}}\U_{\hbar}\ldots\pi_{\gamma_1}\U_{\hbar}\pi_{\gamma_0}\psi_{\hbar},\psi_{\hbar}\rangle
+\sum_{[\gamma]\in \Sigma_n(W_{n_0},\tau)^k}\langle\pi_{\gamma_{N-1}}\U_{\hbar}\ldots\pi_{\gamma_1}\U_{\hbar}\pi_{\gamma_0}\psi_{\hbar},\psi_{\hbar}\rangle
=e^{\frac{\imath (N-1)\overline{z(\hbar)}}{\hbar}}.$$

Using previous lemma, one knows that
$$\left|\sum_{[\gamma]\in (\Sigma_n(W_{n_0},\tau)^k)^c}\langle\pi_{\gamma_{N-1}}\U_{\hbar}\ldots\pi_{\gamma_1}\U_{\hbar}\pi_{\gamma_0}\psi_{\hbar},\psi_{\hbar}\rangle\right|
\leq \theta   \left(1+\mathcal{O}(\hbar^{\nu_0'})\right)^{k}
 e^{\frac{(N-1)\text{Im}(z(\hbar))}{\hbar}}.$$

This allows to derive that
\begin{equation}\label{e:lowboundlaststep}
e^{-\frac{ (kn-1)\text{Im}(z(\hbar))}{\hbar}}\left|\sum_{[\gamma]\in \Sigma_n(W_{n_0},\tau)^k}\langle\pi_{\gamma_{N-1}}\U_{\hbar}\ldots\pi_{\gamma_1}\U_{\hbar}\pi_{\gamma_0}\psi_{\hbar},\psi_{\hbar}\rangle\right|
\geq 1-\theta  \left(1+\mathcal{O}(\hbar^{\nu_0'})\right)^{k}.
\end{equation}

At this point, one can use hyperbolic estimates\footnote{We underline that the Anosov assumption on the geodesic flow is only used at this point of the proof.} on quantum cylinders that were first used in~\cite{An08} and then in several other articles to 
derive quantitative properties of semiclassical measures~\cite{An08, AN07, Riv10a}. Recall that these estimates tell us (Theorem $1.3.3$ in~\cite{An08})
 that for every $\mathcal{K}>0$, there exists $\hbar_{\mathcal{K}}>0$ such that for every $\hbar\leq\hbar_{\mathcal{K}}$, for every $0\leq N\leq \mathcal{K}|\log\hbar|$ 
and for every $[\gamma]$ in $\Sigma_N$,
\begin{equation}\label{e:hypest}
\left\|\Pi_{\gamma}\Op_{\hbar}(\chi_0)\right\|\leq2e^{N\mathcal{O}(\hbar)}(2\pi\hbar)^{-\frac{d}{2}}J^u_N(\gamma)^{\frac{1}{2}}(1+\mathcal{O}(\epsilon))^N,
\end{equation}
where $\epsilon$ is an upper bound on the diameter of our partition, $\mathcal{O}(\hbar)$ depends only on $V$ and on $M$ and the constant in $\mathcal{O}(\epsilon)$ is uniform in $\gamma$ and $\hbar$. 
\begin{rema} 
We underline that the proof of~\eqref{e:hypest} in~\cite{An08} was given in a selfadjoint setting. The fact that we introduce a damping term does not change the proof, one only has to take care that it slightly modifies the WKB expansion. Yet, as $V\geq 0$, one can verify that the exponential decrease due to the damping term can be crudely bounded by a $e^{N\mathcal{O}(\hbar)}$. It explains the apparition of this new term compared with Theorem $1.3.3$ in~\cite{An08}. We refer the reader to~\cite{Sch10} for similar estimates in a nonselfadjoint setting -- see also~\cite{NZ09}. In fact, the estimates from~\cite{Sch10} give a sharper upper bound that we do not need here.
 
\end{rema}

Using inequality~(\ref{e:lowboundlaststep}) and the multiplicative structure of $J^u$, we find that
$$\left(\sum_{[\alpha]\in \Sigma_n(W_{n_0},\tau)}J^{u}_n(\alpha)^{\frac{1}{2}}e^{-\frac{(n-1)\text{Im}(z(\hbar)}{\hbar}}\right)^k\hspace{7cm}$$
$$\geq e^{-k\lambda_0-\frac{(k-1)\text{Im}(z(\hbar)}{\hbar}} \left(1-(1+\mathcal{O}(\hbar^{\nu_0'}))^k\theta
+\mathcal{O}(\hbar^{\infty})\right)\hbar^{d/2}e^{-C_0 k\kappa\epsilon|\log\hbar|},$$
where $C_0$ is a positive constant that depends only on $M$ and $\lambda_0$ is an upper bound on all the $\log J^u(\alpha_0,\alpha_1)$. This concludes the proof of lemma~\ref{l:lowbound}.\\

\subsubsection{Proof of lemma~\ref{l:submultip}} Our proof of lemma~\ref{l:lowbound} relied crucially on lemma~\ref{l:submultip} that we will prove now. We will show that the strategy from~\cite{An08} (proof of lemma $2.2.3$) can be adapted in a nonselfadjoint setting thanks to our assumption on the localization property~(\ref{e:spectralpara}) of $z(\hbar)$.\\

\textbf{First simplification of the sum.}
Let $[\gamma]$ be an element in $(\Sigma_n(W_{n_0},\tau)^k)^c$. It can be decomposed into the concatenation of $n$-cylinders $[\gamma]:=[\gamma^0,\ldots,\gamma^{k-1}]$ 
where at least one of the cylinders $[\gamma^j]$ does not belong to $\Sigma_n(W_{n_0},\tau)$. We write

\begin{equation}\label{e:sumsplit}
\sum_{[\gamma]\in (\Sigma_n(W_{n_0},\tau)^k)^c}\pi_{\gamma_{N-1}}\U_{\hbar}\ldots\pi_{\gamma_1}\U_{\hbar}\pi_{\gamma_0}=
\sum_{j=0}^{k-1}\sum_{(*)}\U_{\hbar}^{-kn+1}\Pi_{\gamma^{k-1}}((k-1)n)\ldots\Pi_{\gamma^{j}}(jn)\ldots\Pi_{\gamma^{0}},
\end{equation}
where $(*)$ means $([\gamma^i]\in \Sigma_n(W_{n_0},\tau)\ \text{for}\ i>j,\ [\gamma^j]\in\Sigma_n(W_{n_0},\tau)^c,\ [\gamma^i]\in\Sigma_n\ \text{for}\ i<j)$. We can use lemma~\ref{l:cutoff} to introduce cutoffs in each term of this sum. Then, using this equality and the property of partition of identity, we find that
\begin{equation}\label{e:crudeupperbound}
\left\|\sum_{[\gamma]\in (\Sigma_n(W_{n_0},\tau)^k)^c}\pi_{\gamma_{N-1}}\U_{\hbar}\ldots\pi_{\gamma_1}\U_{\hbar}\pi_{\gamma_0}\psi_{\hbar}\right\|\leq
\mathbf{R}'\sum_{j=0}^{k-1} \mathbf{R}_{k-j-1}e^{\frac{jn\text{Im}(z(\hbar))}{\hbar}}+\mathcal{O}(\hbar^{\infty}),
\end{equation}
where $$\mathbf{R}':=\left\|\sum_{[\alpha]\in\Sigma_n(W_{n_0},\tau)^c}\pi_{\alpha_{n-1}}\U_{\hbar}\ldots\pi_{\alpha_1}\U_{\hbar}\pi_{\alpha_0}\psi_{\hbar}\right\|_{L^2(M)},$$ 
$\mathbf{R}_{k-1}=1$ and, for $0\leq j\leq k-2$,
$$\mathbf{R}_{k-j-1}:=\prod_{l=j+1}^{k-1}\left\|\sum_{[\alpha]\in\Sigma_n(W_{n_0},\tau)}\pi_{\alpha_{n-1}}\U_{\hbar}\ldots\pi_{\alpha_1}\U_{\hbar}\pi_{\alpha_0}\U_{\hbar}\Op_{\hbar}(\chi_{-(k-j)+1})\right\|.\vspace{1cm}$$

\textbf{Bound on $\mathbf{R}_{k-j-1}$.} Fix now a family $W$ in $\Sigma_n$ and any $0\leq j\leq k-1$. Then, one has
$$\left\|\sum_{[\alpha]\in W}\pi_{\alpha_{n-1}}\ldots\U_{\hbar}\pi_{\alpha_{0}}\U_{\hbar}
\Op_{\hbar}(\chi_{-(k-j)+1})\right\|\leq \left\|\sum_{[\alpha]\in W}\pi_{\alpha_{n-1}}\ldots\U_{\hbar}\pi_{\alpha_{0}}
\U_{\hbar}\Op_{\hbar}(\chi_{-(k-j)+1})\U_{\hbar}^{-n}\right\|\left\|\U_{\hbar}^{n}\right\|.$$
One knows that $\left\|\U_{\hbar}^{n}\right\|\leq 1.$ Moreover, one can verify (see lemma~\ref{l:cylsums} in the appendix) that, for $\kappa$ small enough,
$$\left\|\sum_{[\alpha]\in W}\pi_{\alpha_{n-1}}\ldots\U_{\hbar}\pi_{\alpha_{0}}
\U_{\hbar}\Op_{\hbar}(\chi_{-(k-j)+1})\U_{\hbar}^{-n}\right\|\leq1+\mathcal{O}(\hbar^{\nu_0'}),$$
where $\mathcal{O}(\hbar^{\nu_0'})$ depends on the choice of the partition (and on $P_0$) and one has $\nu_0'>0$. Moreover, the constant in the remainder is uniform for $W\subset \Sigma_n$. Finally, one has, for any $0\leq j\leq k-1$, 
$$\mathbf{R}_{k-j-1}\leq\left((1+\mathcal{O}(\hbar^{\nu_0'}))\right)^{k-j-1}.$$

\textbf{The conclusion.} We now implement these upper bounds in inequality~\eqref{e:crudeupperbound} and we find that
$$\left\|\sum_{[\gamma]\in (\Sigma_n(W_{n_0},\tau)^k)^c}\pi_{\gamma_{N-1}}\U_{\hbar}\ldots\pi_{\gamma_1}\U_{\hbar}\pi_{\gamma_0}\psi_{\hbar}\right\|\hspace{4cm}$$
$$\leq
\sum_{j=0}^{k-1} \left((1+\mathcal{O}(\hbar^{\nu_0'}))\right)^{k-j-1}e^{\frac{jn\text{Im}(z(\hbar))}{\hbar}}
\left\|\sum_{[\alpha]\in\Sigma_n(W_{n_0},\tau)^c}\pi_{\alpha_{n-1}}\U_{\hbar}\ldots\pi_{\alpha_1}\U_{\hbar}\pi_{\alpha_0}\psi_{\hbar}\right\|_{L^2(M)}.$$
Then, thanks to our assumption on $\|\sum_{[\alpha]\in\Sigma_n(W_{n_0},\tau)^c}
\pi_{\alpha_{n-1}}\U_{\hbar}\ldots\pi_{\alpha_1}\U_{\hbar}\pi_{\alpha_0}\psi_{\hbar}\|_{L^2(M)}$, we find
$$\left\|\sum_{[\gamma]\in (\Sigma_n(W_{n_0},\tau)^k)^c}\pi_{\gamma_{N-1}}\U_{\hbar}\ldots\pi_{\gamma_1}\U_{\hbar}\pi_{\gamma_0}\psi_{\hbar}\right\|\leq
 \frac{\theta e^{-Ck\kappa}}{k}\sum_{j=0}^{k-1} \left((1+\mathcal{O}(\hbar^{\nu_0'}))e^{n\mathcal{O}(\hbar)}\right)^{k-j-1}
e^{\frac{((j+1)n-1)\text{Im}(z(\hbar))}{\hbar}}.$$
Recall now that we made the assumption that
$$\frac{\text{Im}(z(\hbar))}{\hbar}\geq -C|\log\hbar|^{-1}.$$ 
This assumption is crucial because it allows to bound $ 
e^{\frac{((j+1)n-1)\text{Im}(z(\hbar))}{\hbar}}$ by $e^{\frac{(kn-1)\text{Im}(z(\hbar))}{\hbar}}e^{Ck\kappa}$ for every $0\leq j\leq k-1$. It implies
$$\left\|\sum_{[\gamma]\in (\Sigma_n(W_{n_0},\tau)^k)^c}\pi_{\gamma_{N-1}}\U_{\hbar}\ldots\pi_{\gamma_1}\U_{\hbar}\pi_{\gamma_0}\psi_{\hbar}\right\|\leq
 \theta  \left(1+\mathcal{O}(\hbar^{\nu_0'})\right)^{k}
 e^{\frac{(kn-1)\text{Im}(z(\hbar))}{\hbar}},$$
that concludes the proof of lemma~\ref{l:submultip}.

\subsection{Subinvariance of the quantum functional $\mu_{\hbar}^{\Sigma}$}
\label{ss:subinvariance}

 In paragraph~\ref{ss:partition}, we constructed a quantum functional $\mu_{\hbar}^{\Sigma}$ on a set $\Sigma$. The set $\Sigma$ was defined from a regularized partition that we will suppose to be fixed in this section. As mentioned in paragraph~\ref{ss:partition}, this functional is not invariant under the shift $\sigma$ and it only satisfies
$$\forall [\alpha_0,\ldots,\alpha_{n_1-1}],\ \forall p\geq 0,\ \mu_{\hbar}^{\Sigma}(\sigma^{-p}[\alpha_0,\ldots,\alpha_{n_1-1}])=\mu_{\hbar}^{\Sigma}([\alpha_0,\ldots,\alpha_{n_1-1}])+o_{n_1,p}(1).$$
This remainder term in this property is not explicit enough to use it directly. Yet, as mentioned above, one can use the fact that the $(\psi_{\hbar})_{\hbar}$ are eigenmodes to write
\begin{equation}\label{e:subinv}\mu_{\hbar}^{\Sigma}([\alpha_0,\ldots,\alpha_{n_1-1}])=e^{-\frac{2p\text{Im}(z(\hbar))}{\hbar}}\sum_{|\beta|=p}\left\langle(\U_{\hbar}^p)^*\U_{\hbar}^p\Pi_{\beta.\alpha}\psi_{\hbar},\psi_{\hbar}\right\rangle.\end{equation}
Starting from this observation, we will prove the following subinvariance property:
\begin{prop}\label{p:subinvariance} Let $(P_i^{\hbar})_{i=1,\ldots K}$ be a fixed partition satisfying the assumptions from paragraph~\ref{s:discretization}. There exist\footnote{Even if we will not mention it at every step of the proof, both $\kappa_0$ and $\nu_0$ depend on the partition $\mathcal{M}$, on the energy layer we work on and on the parameter $\overline{\nu}$  used for the smoothing of the partition.} $\kappa_0>0$ and $\nu_0$ such that for every $0\leq p\leq \kappa_0|\log\hbar|$, one has 
$$\forall [\alpha_0,\ldots,\alpha_{n_1-1}],\ \mu_{\hbar}^{\Sigma}([\alpha_0,\ldots,\alpha_{n_1-1}])\leq e^{-\frac{2p\text{Im}(z(\hbar))}{\hbar}}\left(\mu_{\hbar}^{\Sigma}(\sigma^{-p}[\alpha_0,\ldots,\alpha_{n_1-1}])+K^p\mathcal{O}_{n_1}(\hbar^{\nu_0})\right).$$
 \end{prop}

Before giving the proof of this proposition, let us mention that once the partition is fixed (hence $K$ is fixed), there exists a $\kappa_1>0$ such that for every $\kappa\leq\kappa_1$, the remainder $K^p\mathcal{O}_{n_1}(\hbar^{\nu_0})$ in the proposition is of order $\mathcal{O}_{n_1}(\hbar^{\nu'(\kappa)})$ for some positive $\nu'(\kappa)$.

\subsubsection*{Proof of proposition~\ref{p:subinvariance}} The proof of proposition~\ref{p:subinvariance} is a direct consequence of equality~(\ref{e:subinv}) and of the following lemma:

\begin{lemm}\label{l:subinvariance} Let $(P_i^{\hbar})_{i=1,\ldots K}$ be a fixed partition satisfying the assumptions from paragraph~\ref{s:discretization}. There exists $\kappa_0>0$ and $\nu_0$ such that for every $0\leq p\leq \kappa_0|\log\hbar|$, one has 
$$\forall [\alpha]\in\Sigma_{n_1},\ \forall [\beta]\in\Sigma_{p},\ \left\langle(\U_{\hbar}^p)^*\U_{\hbar}^p\Pi_{\beta.\alpha}\psi_{\hbar},\psi_{\hbar}\right\rangle\leq \left(\left\langle\Pi_{\beta.\alpha}\psi_{\hbar},\psi_{\hbar}\right\rangle+\mathcal{O}_{n_1}(\hbar^{\nu_0})\right).$$

\end{lemm}

In order to prove lemma~\ref{l:subinvariance}, we first use lemma~\ref{l:productPDO} from the appendix and the fact that $\Op_{\hbar}(\chi_0)\psi_{\hbar}=\psi_{\hbar}+\mathcal{O}(\hbar^{\infty})$. It allows us to write, for $0\leq p\leq\kappa_0|\log\hbar|$,
$$\Pi_{\beta.\alpha}\psi_{\hbar}=\Op_{\hbar}\left(P_{\beta_0}\ldots P_{\beta_{p-1}}\circ g^{p-1} P_{\alpha_{p}}\circ g^p\ldots P_{\alpha_{n_1+p-1}}\circ g^{n_1+p-1}\chi_0\right)\psi_{\hbar}+\mathcal{O}_{n_1}(\hbar^{\nu_0}),$$
where $\nu_0>0$ and $\kappa_0$ are given by the lemmas in the appendix. Then, one can use Egorov property (see paragraph~\ref{pa:egorov}) to write that, for every $0\leq p\leq\kappa_0|\log\hbar|$,
$$(\U_{\hbar}^p)^*\U_{\hbar}^p\Pi_{\beta.\alpha}\psi_{\hbar}\hspace{12cm}$$
$$\hspace{0,5cm}=\Op_{\hbar}\left(e^{-2\int_0^pV\circ g^sds}P_{\beta_0}\ldots P_{\beta_{p-1}}\circ g^{p-1} P_{\alpha_{p}}\circ g^p\ldots P_{\alpha_{n_1+p-1}}\circ g^{n_1+p-1}\chi_0\right)\psi_{\hbar}+\mathcal{O}_{n_1}(\hbar^{\nu_0}),$$
where $\nu_0>0$ is still a positive constant. Thanks to lemma~\ref{l:productPDO} from the appendix, one knows that $e^{-2\int_0^pV\circ g^sds}P_{\beta_0}\ldots P_{\beta_{p-1}}\circ g^{p-1} P_{\alpha_{p}}\circ g^p\ldots P_{\alpha_{n_1+p-1}}\circ g^{n_1+p-1}\chi_0$ belongs to a good class of symbols $S^{-\infty,0}_{\overline{\nu}_0}(T^*M)$ (for some $0\leq\overline{\nu}_0<1/2$). Moreover, it is compactly supported in $T^*M$; hence, it can also be quantized using $\Op_{\hbar}^+$ (see paragraph~\ref{p:antiwick}). It implies that
$$(\U_{\hbar}^p)^*\U_{\hbar}^p\Pi_{\beta.\alpha}\psi_{\hbar}\hspace{12cm}$$
$$\hspace{0,5cm}=\Op_{\hbar}^+\left(e^{-2\int_0^pV\circ g^sds}P_{\beta_0}\ldots P_{\beta_{p-1}}\circ g^{p-1} P_{\alpha_{p}}\circ g^p\ldots P_{\alpha_{n_1+p-1}}\circ g^{n_1+p-1}\chi_0\right)\psi_{\hbar}+\mathcal{O}_{n_1}(\hbar^{\nu_0}),$$
where $\nu_0>0$. Thanks to the positivity of $\Op_{\hbar}^+$, one finds then
$$\langle(\U_{\hbar}^p)^*\U_{\hbar}^p\Pi_{\beta.\alpha}\psi_{\hbar},\psi_{\hbar}\rangle \hspace{11cm}$$
$$\leq \langle\Op_{\hbar}^+\left(P_{\beta_0}\ldots P_{\beta_{p-1}}\circ g^{p-1} P_{\alpha_{p}}\circ g^p\ldots P_{\alpha_{n_1+p-1}}\circ g^{n_1+p-1}\chi_0\right)\psi_{\hbar},\psi_{\hbar}\rangle+\mathcal{O}_{n_1}(\hbar^{\nu_0}).$$
Using finally the same arguments backward, one finds
$$\langle(\U_{\hbar}^p)^*\U_{\hbar}^p\Pi_{\beta.\alpha}\psi_{\hbar},\psi_{\hbar}\rangle\leq \langle\Pi_{\beta.\alpha}\psi_{\hbar},\psi_{\hbar}\rangle+\mathcal{O}_{n_1}(\hbar^{\nu_0}),$$
where $\nu_0>0$ still satisfies the same properties as above.

\subsection{Using the fact that $(P_j^{\hbar})_{j=1,\ldots,K}$ are almost orthogonal}\label{p:semiclassicalapprox} In this last paragraph, we will prove that, for $\kappa$ small enough,
$$\left\|\sum_{[\alpha]\in\Sigma_n(W_{n_0},\tau)^c}\pi_{\alpha_{n-1}}\ldots\pi_{\alpha_1}(2-n)\pi_{\alpha_0}(1-n)\psi_{\hbar}\right\|_{L^2(M)}^2\hspace{3cm}$$
$$=\sum_{[\alpha]\in\Sigma_n(W_{n_0},\tau)^c}\left\langle\pi_{\alpha_{n-1}}\ldots\pi_{\alpha_1}(2-n)\pi_{\alpha_0}(1-n)\psi_{\hbar},\psi_{\hbar}\right\rangle
+\mathcal{O}(\hbar^{\nu_0'}),$$
where $\nu_0'$ is a positive constant that depends on the choice of the partition, on the choice of the energy layer and on the parameter $\overline{\nu}$ and $b$ used for the smoothing of the partition. The proof of this property relies on the fact that for $\kappa$ small enough, cylinders of length $n=[\kappa|\log\hbar|]$ are amenable to semiclassical rules and that $(P_j^{\hbar})_{j=1,\ldots, K}$ acts almost like a family of orthogonal projectors on $\psi_{\hbar}$.

The proof of this equality was already given in~\cite{An08} in the case $V\equiv0$. If proceeding carefully, the arguments can be adapted in our setting and for the sake of completeness, we give a proof of this equality below.

For simplicity of notations, we introduce $\mathbf{P}_{\alpha}^{\hbar}=P_{\alpha_{n-1}}^{\hbar}\ldots P_{\alpha_1}^{\hbar}\circ g^{2-n}P_{\alpha_0}^{\hbar}\circ g^{1-n}$. As in lemma~\ref{l:productPDO}, one can prove that, for $0\leq\kappa\leq\kappa_0$,
$$\left\|\sum_{[\alpha]\in\Sigma_n(W_{n_0},\tau)^c}\pi_{\alpha_{n-1}}\ldots\pi_{\alpha_1}(2-n)\pi_{\alpha_0}(1-n)\psi_{\hbar}\right\|_{L^2(M)}^2\hspace{3cm}$$
$$=\sum_{[\alpha], [\alpha']\in\Sigma_n(W_{n_0},\tau)^c}\left\langle\Op_{\hbar}\left(\mathbf{P}_{\alpha}^{\hbar}\chi_0\right)\psi_{\hbar},\Op_{\hbar}\left(\mathbf{P}_{\alpha'}^{\hbar}\chi_0\right)\psi_{\hbar}\right\rangle
+(\sharp\Sigma_n)^2\mathcal{O}(\hbar^{\nu_0}),$$
where we used the fact that $\Op_{\hbar}(\chi_0)\psi_{\hbar}=\psi_{\hbar}+\mathcal{O}(\hbar^{\infty})$ and where $\nu_0>0$. The parameter $\nu_0$ also depends on the choice of the partition, on the choice of the energy layer and 
on the regularization parameter $\overline{\nu}$ used for the smoothing of the partition. We will omit to mention this dependence in the following of the proof and we will also allow to take $\nu_0>0$ to be smaller to have the semiclassical arguments below work. 

We underline that $\mathbf{P}_{\alpha}^{\hbar}\chi_0$ belongs to a class of symbols of type $S^{-\infty,0}_{\overline{\nu}_0}(T^*M)$ (where $0\leq\overline{\nu}_0<1/2$). Thanks to composition rules for pseudodifferential operators, we derive that
$$\left\|\sum_{[\alpha]\in\Sigma_n(W_{n_0},\tau)^c}\pi_{\alpha_{n-1}}\ldots\pi_{\alpha_1}(2-n)\pi_{\alpha_0}(1-n)\psi_{\hbar}\right\|_{L^2(M)}^2\hspace{3cm}$$
\begin{equation}\label{e:steporthproj}=\sum_{[\alpha], [\alpha']\in\Sigma_n(W_{n_0},\tau)^c}\left\langle\Op_{\hbar}\left(\mathbf{P}_{\alpha'}^{\hbar}\mathbf{P}_{\alpha}^{\hbar}\chi_0^2\right)\psi_{\hbar},\psi_{\hbar}\right\rangle
+(\sharp\Sigma_n)^2\mathcal{O}(\hbar^{\nu_0}).\end{equation}
We will now distinguish two kind of terms.

\subsubsection*{First case $\alpha=\alpha'$} We use the composition formula and the long time Egorov property to derive that
$$\left\|\Op_{\hbar}\left((\mathbf{P}_{\alpha}^{\hbar})^2\chi_0^2\right)-\Op_{\hbar}\left(\frac{(\mathbf{P}_{\alpha}^{\hbar})^2}{(P_{\alpha_{0}}^{\hbar}\circ g^{1-n})^2}\chi_0^2\right)\pi_{\alpha_{0}}^2(1-n)\right\|_{L^2\rightarrow L^2}=\mathcal{O}(\hbar^{\nu_0}),$$
where $\nu_0>0$ and the remainder can be chosen uniform in $\alpha$. Then, one can use the specific properties of our partition -- precisely, the fact that it behaves like orthogonal projectors when it acts on $\psi_{\hbar}$ (see paragraph~\ref{s:discretization}). It allows to prove that
$$\left\|\left(\pi_{\alpha_{0}}(1-n)-\pi_{\alpha_{0}}^2(1-n)\right)\psi_{\hbar}\right\|=\mathcal{O}(\hbar^{\frac{b}{2}}).$$
Using one more time the composition formula and the long time Egorov property, we find that
$$\left\|\Op_{\hbar}\left((\mathbf{P}_{\alpha}^{\hbar})^2\chi_0^2\right)\psi_{\hbar}-Op_{\hbar}\left(\frac{(\mathbf{P}_{\alpha}^{\hbar})^2}{P_{\alpha_{0}}^{\hbar}\circ g^{1-n}}\chi_0^2\right)\psi_{\hbar}\right\|=\mathcal{O}(\hbar^{\nu_0})+\mathcal{O}(\hbar^{\frac{b}{2}}),$$
where the constant in the remainder is still uniform in $\alpha$. Proceeding by induction, we get the following approximation:  
$$\left\|\Op_{\hbar}\left((\mathbf{P}_{\alpha}^{\hbar})^2\chi_0^2\right)\psi_{\hbar}-Op_{\hbar}\left(\mathbf{P}_{\alpha}^{\hbar}\chi_0^2\right)\psi_{\hbar}\right\|=|\log\hbar|\left(\mathcal{O}(\hbar^{\nu_0})+\mathcal{O}(\hbar^{\frac{b}{2}})\right).$$
Finally, thanks to lemma~\ref{l:productPDO}, we obtain
$$\left\|\Op_{\hbar}\left((\mathbf{P}_{\alpha}^{\hbar})^2\chi_0^2\right)\psi_{\hbar}-\pi_{\alpha_{n-1}}\ldots\pi_{\alpha_1}(2-n)\pi_{\alpha_0}(1-n)\psi_{\hbar}\right\|=|\log\hbar|\left(\mathcal{O}(\hbar^{\nu_0})+\mathcal{O}(\hbar^{\frac{b}{2}})\right),$$
with an uniform constant in $\alpha$ in the remainder.

\subsubsection*{Second case $\alpha\neq\alpha'$} In this case, there exists $j$ such that $\alpha_j\neq\alpha_j'$. As in the first case, we use the composition formula and the Egorov property to write
$$\left\|\Op_{\hbar}\left(\mathbf{P}_{\alpha}^{\hbar}\mathbf{P}_{\alpha'}^{\hbar}\chi_0^2\right)-\Op_{\hbar}\left(\frac{\mathbf{P}_{\alpha}^{\hbar}\mathbf{P}_{\alpha'}^{\hbar}}{P_{\alpha_{j}}^{\hbar}\circ g^{1-n}\times P_{\alpha_{j}'}^{\hbar}\circ g^{1-n}}\chi_0^2\right)(\pi_{\alpha_{j}}\pi_{\alpha_{j}'})(1-n)\right\|_{L^2\rightarrow L^2}=\mathcal{O}(\hbar^{\nu_0}),$$
with the same properties as above for the remainder. Recall again that our partition behaves like orthogonal projectors when it acts on $\psi_{\hbar}$ (see paragraph~\ref{s:discretization}). Hence, one finds
$$\left\|(\pi_{\alpha_{j}}\pi_{\alpha_{j}'})(1-n)\psi_{\hbar}\right\|=\mathcal{O}(e^{2\kappa\|V\|_{\infty}|\log\hbar|}\hbar^{\frac{b}{2}}).$$
Thanks to the Calder\'on-Vaillancourt Theorem, the operator $\displaystyle\Op_{\hbar}\left(\frac{\mathbf{P}_{\alpha}^{\hbar}\mathbf{P}_{\alpha'}^{\hbar}}{P_{\alpha_{j}}^{\hbar}\circ g^{1-n}\times P_{\alpha_{j}'}^{\hbar}\circ g^{1-n}}\chi_0^2\right)$ has a norm bounded by a constant uniform in $\alpha$ and in $\hbar$. Finally, we obtain
$$\left\|\Op_{\hbar}\left(\mathbf{P}_{\alpha}^{\hbar}\mathbf{P}_{\alpha'}^{\hbar}\chi_0^2\right)\psi_{\hbar}\right\|=\mathcal{O}(\hbar^{\nu_0})+\mathcal{O}(e^{2\kappa\|V\|_{\infty}|\log\hbar|}\hbar^{\frac{b}{2}}).$$

\subsubsection*{Combining the two cases with~\eqref{e:steporthproj}} To conclude the proof of this paragraph, we combine equality~\eqref{e:steporthproj} with the two cases treated above. We find that, for $\kappa$ small enough, there exists a constant $\nu_0>0$ (depending also on the partition $\mathcal{M}$, the size of the energy layer and the smoothing parameters $\overline{\nu}$ and $b$) and such that
$$\left\|\sum_{[\alpha]\in\Sigma_n(W_{n_0},\tau)^c}\pi_{\alpha_{n-1}}\ldots\pi_{\alpha_1}(2-n)\pi_{\alpha_0}(1-n)\psi_{\hbar}\right\|_{L^2(M)}^2\hspace{4cm}$$
$$=\sum_{[\alpha]\in\Sigma_n(W_{n_0},\tau)^c}\left\langle\pi_{\alpha_{n-1}}\ldots\pi_{\alpha_1}(2-n)\pi_{\alpha_0}(1-n)\psi_{\hbar},\psi_{\hbar}\right\rangle
+(\sharp\Sigma_n)^2\mathcal{O}(\hbar^{\nu_0}).$$
As $\sharp\Sigma_n$ is equal to $K^n$ as $\nu_0>0$ can be chosen uniformly for $\kappa$ small enough, one can find $\kappa$ small enough to have a remainder which goes to $0$ as a positive power of $\hbar$ (which was the expected property).

\appendix

\section{Pseudodifferential calculus on a manifold}

In this appendix, we review some basic facts on semiclassical analysis that can be found for instance in~\cite{DS, EZ}. We also give several lemmas that we use at different steps of the paper.

\subsection{General facts}

\label{a:pdo}

Recall that we define on $\mathbb{R}^{2d}$ the following class of symbols:
$$S^{m,k}(\mathbb{R}^{2d}):=\left\{(a_{\hbar}(x,\xi))_{\hbar\in(0,1]}\in C^{\infty}(\mathbb{R}^{2d}):|\partial^{\alpha}_x\partial^{\beta}_{\xi}a_{\hbar}|
\leq C_{\alpha,\beta}\hbar^{-k}\langle\xi\rangle^{m-|\beta|}\right\}.$$
Let $M$ be a smooth Riemannian $d$-manifold without boundary. Consider a smooth atlas $(f_l,V_l)$ of $M$, where each $f_l$ is a smooth diffeomorphism from 
$V_l\subset M$ to a bounded open set $W_l\subset\mathbb{R}^{d}$. To each $f_l$ correspond a pull back $f_l^*:C^{\infty}(W_l)\rightarrow C^{\infty}(V_l)$ and a canonical 
map $\tilde{f}_l$ from $T^*V_l$ to $T^*W_l$:
$$\tilde{f}_l:(x,\xi)\mapsto\left(f_l(x),(Df_l(x)^{-1})^T\xi\right).$$
Consider now a smooth locally finite partition of identity $(\phi_l)$ adapted to the previous atlas $(f_l,V_l)$. 
That means $\sum_l\phi_l=1$ and $\phi_l\in C^{\infty}(V_l)$. Then, any observable $a$ in $C^{\infty}(T^*M)$ can be decomposed as follows: $a=\sum_l a_l$, where 
$a_l=a\phi_l$. Each $a_l$ belongs to $C^{\infty}(T^*V_l)$ and can be pushed to a function $\tilde{a}_l=(\tilde{f}_l^{-1})^*a_l\in C^{\infty}(T^*W_l)$. 
As in~\cite{DS, EZ}, define the class of symbols of order $m$ and index $k$
\begin{equation}
\label{defpdo}S^{m,k}(T^{*}M):=\left\{a_{\hbar}\in C^{\infty}(T^*M\times(0,1]):|\partial^{\alpha}_x\partial^{\beta}_{\xi}a_{\hbar}|\leq C_{\alpha,\beta}\hbar^{-k}\langle\xi\rangle^{m-|\beta|}\right\}.
\end{equation}
Then, for $a\in S^{m,k}(T^{*}M)$ and for each $l$, one can associate to the symbol $\tilde{a}_l\in S^{m,k}(\mathbb{R}^{2d})$ the standard Weyl quantization
$$\Op_{\hbar}^{w}(\tilde{a}_l)u(x):=
\frac{1}{(2\pi\hbar)^d}\int_{R^{2d}}e^{\frac{\imath}{\hbar}\langle x-y,\xi\rangle}\tilde{a}_l\left(\frac{x+y}{2},\xi;\hbar\right)u(y)dyd\xi,$$
where $u\in\mathcal{S}(\mathbb{R}^d)$, the Schwartz class. Consider now a smooth cutoff $\psi_l\in C_c^{\infty}(V_l)$ such that $\psi_l=1$ close to the support of $\phi_l$. 
A quantization of $a\in S^{m,k}$ is then defined in the following way:
\begin{equation}
\label{pdomanifold}\Op_{\hbar}(a)(u):=\sum_l \psi_l\times\left(f_l^*\Op_{\hbar}^w(\tilde{a}_l)(f_l^{-1})^*\right)\left(\psi_l\times u\right),
\end{equation}
where $u\in C^{\infty}(M)$. This quantization procedure $\Op_{\hbar}$ sends (modulo $\mathcal{O}(\hbar^{\infty})$) $S^{m,k}(T^{*}M)$ onto the space of pseudodifferential 
operators of order $m$ and of index $k$, denoted $\Psi^{m,k}(M)$~\cite{DS, EZ}. It can be shown that the dependence in the cutoffs $\phi_l$ and $\psi_l$ only appears at order 
$2$ in $\hbar$ (using for instance theorem $18.1.17$ in~\cite{Ho}) and the principal symbol map $\sigma_0:\Psi^{m,k}(M)\rightarrow S^{m,k}/S^{m,k-1}(T^{*}M)$ is then 
intrinsically defined. Most of the rules (for example the composition of operators, the Egorov and Calder\'on-Vaillancourt Theorems) that holds in the case of 
$\mathbb{R}^{2d}$ still holds in the case of $\Psi^{m,k}(M)$. Because our study concerns behavior of quantum evolution for logarithmic times in $\hbar$, a larger class of 
symbols should be introduced as in~\cite{DS, EZ}, for $0\leq\overline{\nu}<1/2$,
\begin{equation}\label{symbol}
S^{m,k}_{\overline{\nu}}(T^{*}M):=\left\{a_{\hbar}\in C^{\infty}(T^*M\times(0,1]):
|\partial^{\alpha}_x\partial^{\beta}_{\xi}a_{\hbar}|\leq C_{\alpha,\beta}\hbar^{-k-\overline{\nu}|\alpha+\beta|}\langle\xi\rangle^{m-|\beta|}\right\}.
\end{equation}
Results of~\cite{DS, EZ} can be applied to this new class of symbols. For example, a symbol of $S^{0,0}_{\overline{\nu}}(T^*M)$ gives a bounded operator on $L^2(M)$ 
(with norm uniformly bounded with respect to $\hbar$).

\subsection{Positive quantization}\label{p:antiwick}

Even if the Weyl procedure is a natural choice to quantize an observable $a$ on $\mathbb{R}^{2d}$, it is sometimes preferrable to use a quantization procedure $\Op_{\hbar}$ that satisfies the property~: $\Op_{\hbar}(a)\geq 0$ if $a\geq0$. This can be achieved thanks to the anti-Wick procedure $\Op_{\hbar}^{AW}$, see~\cite{HeMaRo}.
 For $a$ in $S^{0,0}_{\overline{\nu}}(\mathbb{R}^{2d})$, that coincides with a function on $\mathbb{R}^d$ outside a compact subset of $T^*\mathbb{R}^d=\mathbb{R}^{2d}$, one has
\begin{equation}\label{equivalence-positive-quantization}\|\Op_{\hbar}^w(a)-\Op_{\hbar}^{AW}(a)\|_{L^2}\leq C\sum_{|\alpha|\leq D}\hbar^{\frac{|\alpha|+1}{2}}\|\partial^{\alpha}da\|,
\end{equation}
where $C$ and $D$ are some positive constants that depend only on the dimension $d$.
To get a positive procedure of quantization on a manifold, one can replace the Weyl quantization by the anti-Wick one in definition~(\ref{pdomanifold}). We will denote $\Op_{\hbar}^+(a)$ this new choice of quantization, well defined for every element in $S^{0,0}_{\overline{\nu}}(T^*M)$ of the form $c_0(x)+c(x,\xi)$ where $c_0$ belongs to $S^{0,0}_{\overline{\nu}}(T^*M)$ and $c$ belongs to $\mathcal{C}^{\infty}_o(T^*M)\cap S^{0,0}_{\overline{\nu}}(T^*M)$. 

This positivity assumption was used at several steps of the paper when we argued that the functional $\mu_{\hbar}^{\Sigma}$ was ``almost positive'' (see for instance paragraphs~\ref{ss:strategy} and~\ref{p:conclusion}) or when we proved that it was ``subinvariant'' (paragraph~\ref{ss:subinvariance}). 


\subsection{Egorov property for long times}\label{pa:egorov}

In this paragraph, we prove an Egorov property for times of order $\kappa|\log\hbar|$, where $\kappa$ is a small enough constant. Consider $q_1$ and $q_2$ two smooth 
functions on $T^*M$ belonging to $S^{0,0}(T^*M)$. We consider a smooth function $a$ on $T^*M$ which is compactly supported in a neighborhood of $S^*M$, say 
$\text{supp}(a)\subset\{(x,\xi):\|\xi\|^2\in[1/2,3/2]\}$ and which belongs to $S^{-\infty,0}_{\nu}$ (with $\nu<1/2)$) We would like to prove that the following operator is a pseudodifferential operator, for every $t\geq 0$,
$$B(t,a)=\left(e^{-\frac{\imath t}{\hbar}\left(-\frac{\hbar^2\Delta}{2}-\imath\hbar\Op_{\hbar}(q_1)\right)}\right)^*\Op_{\hbar}(a)
e^{-\frac{\imath t}{\hbar}\left(-\frac{\hbar^2\Delta}{2}-\imath\hbar\Op_{\hbar}(q_2)\right)}.$$
We underline that the selfadjoint parts of the two quantum propagators are identical while the selfadjoint part are different. We will give two applivations of this 
result at the end of the paragraph: one where $q_1=-q_2$ and one where $q_1=0$. Our proof is taken from~\cite{Roy10} (section $3.3$, Theorem $3.43$) and follows the classical
proof of the Egorov Theorem~\cite{DS, EZ, BoRo02}. Denote $q=\overline{q_1}+q_2$ and 
introduce, for $t\in\mathbb{R}$
$$a_t(s):=a\circ g^{t-s}\exp\left(-\int_0^{t-s}q\circ g^{\tau}d\tau\right).$$
For simplicity of notations, one can use the notation
$$\U_{\hbar}^s(q_1):=e^{-\frac{\imath s}{\hbar}\left(-\frac{\hbar^2\Delta}{2}-\imath\hbar\Op_{\hbar}(q_1)\right)}\ \text{and}\ \U_{\hbar}^s(q_2):=e^{-\frac{\imath s}{\hbar}\left(-\frac{\hbar^2\Delta}{2}-\imath\hbar\Op_{\hbar}(q_2)\right)}.$$
We also introduce the operator
$$R(\hbar,s)=\left(\U_{\hbar}^s(q_1)\right)^*\Op_{\hbar}(a_t(s))\U_{\hbar}^s(q_2).$$
As in the classical proof of the Egorov theorem (i.e. in the selfadjoint case), one can compute the derivative of $R(\hbar,s)$ and finds

$$\frac{d}{ds}\left(R(\hbar,s)\right)=
(\mathcal{U}^s_{\hbar}(q_1))^*\left(\frac{\imath}{\hbar}\left[-\frac{\hbar^2\Delta}{2},\Op_{\hbar}(a_t(s))\right]-\Op_{\hbar}(q_1)^*\Op_{\hbar}(a(s))-
\Op_{\hbar}(a_t(s))\Op_{\hbar}(q_2)\right)\mathcal{U}^s_{\hbar}(q_2)$$
$$\hspace{4cm}-(\mathcal{U}^s_{\hbar}(q_1))^*\left(\Op_{\hbar}\left(\{H,a_t(s)\}\right)-\Op_{\hbar}(a(s)(\overline{q_1}+q_2))\right)\mathcal{U}^s_{\hbar}(q_2).$$
We integrate this equality between $0$ and $t$. Using the standard rules for pseudodifferential calculus (perfomed locally on each chart)~\cite{DS, EZ} (respectively Chapter $7$ and $4$) and the fact 
that $\mathcal{U}^s_{\hbar}(q_2)$ is a bounded operator (with a norm depending\footnote{It is in fact bounded by a constant of order $e^{|s|\|q_2\|_{\infty}}$.} on $q_2$ and $s$), one can then finds that 
$\left(\U_{\hbar}^t(q_1)\right)^*\Op_{\hbar}(a)\U_{\hbar}^t(q_2)$ is a pseudo differential operator in $\Psi^{-\infty,0}(M)$ and modulo $\mathcal{O}(\hbar^{\infty})$, it can 
be written as
$$\left(\U_{\hbar}^t(q_1)\right)^*\Op_{\hbar}(a)\U_{\hbar}^t(q_2)=\Op_{\hbar}(\tilde{a}(t))+\mathcal{O}(\hbar^{\infty}),$$
where $\tilde{a}(t)\sim\sum_{j\geq 0}\hbar^ja_j(t)$, 
$$a_0(t)=a\circ g^{t}\exp\left(-\int_0^{t}(\overline{q_1}+q_2)\circ g^{\tau}d\tau\right),$$ 
and all the other terms $(a_j(t))_{j\geq1}$ in the asymptotic 
expansion depends on $a$, $t$, $q_1$, $q_2$ and the choice of coordinates on the manifold. Moreover, for a fixed $t\in\mathbb{R}$, one can verify that every term in the 
asymptotic expansion has a compact support included in $g^{-t}\text{supp}(a)$ and can be written as $b_j(t)\exp\left(-\int_0^{t}(\overline{q_1}+q_2)\circ g^{\tau}d\tau\right)$ where $b_j(t)$ belongs to $S^{-\infty,0}(T^*M)$. 
In particular, Calder\'on-Vaillancourt Theorem ~\cite{EZ} (chapter $5$) tells us that one can extract constants $C_{a,t}$ and $C_{a,t}'$ (depending on $a$, $q_1$, $q_2$, $t$ and $M$) 
such that
$$
 \left\|\left(\U_{\hbar}^t(q_1)\right)^*\Op_{\hbar}(a)\U_{\hbar}^t(q_2)\right\|_{L^2(M)\rightarrow L^2(M)}\leq C_{a,t}\|a_0(t)\|_{\infty},$$
and also
\begin{equation}\label{e:generalegorov}
 \left\|\left(\U_{\hbar}^t(q_1)\right)^*\Op_{\hbar}(a)\U_{\hbar}^t(q_2)-\Op_{\hbar}(a_0(t))\right\|_{L^2(M)\rightarrow L^2(M)}\leq C_{a,t}'\hbar.\end{equation}

All this discussion was done for fixed $t\geq 0$. In the proof of our main theorems, we needed to apply Egorov property for long range of times of order $\kappa|\log\hbar|$~\cite{BoRo02}. In fact, all the arguments above can be adapted if we use more general class of symbols, i.e. $S_{\overline{\nu}'}^{-\infty,0}(T^*M)$ (see ~\cite{AN07}-section 
$5.2$) with $\overline{\nu}<\overline{\nu}'<1/2$. In particular, one can prove the following uniform estimates:
\begin{prop}\label{p:egorov} There exist constants $\kappa_0>0$ and $\nu_0$ (depending only on $q_1$, $q_2$, $\overline{\nu}$ and the manifold) such that for every smooth function $a$ compactly supported in $\{(x,\xi):\|\xi\|^2\in[1/2,3/2]\}$ and belonging to $S^{-\infty,0}_{\overline{\nu}}(T^*M)$ (with $\overline{\nu}<1/2)$), there exists a constant $C_a>0$ such that 
for every $-\kappa_0|\log\hbar|\leq t\leq \kappa_0|\log\hbar|$, one has
$$ \left\|\left(\U_{\hbar}^t(q_1)\right)^*\Op_{\hbar}(a)\U_{\hbar}^t(q_2)-\Op_{\hbar}(a_0(t))\right\|_{L^2(M)\rightarrow L^2(M)}\leq C_{a}\hbar^{\nu_0}.$$
\end{prop}

\begin{rema}
In this article, we mainly use these Egorov properties in two situations. The first one is when $q_1=q_2=\sqrt{2z(\hbar)}V$. With the notations of the introduction, it means considering the operator $\left(\U_{\hbar}^t\right)^*\Op_{\hbar}(a)\U_{\hbar}^t$. In this case, the principal symbol is $a\circ g^{t}\exp\left(-2\int_0^{t}V\circ g^{\tau}d\tau\right)$.The second situation is when $q_1=\overline{\sqrt{2z(\hbar)}}V$ and $q_2=-\sqrt{2z(\hbar)}V$. With the notations of the introduction, it means considering the operator $\U_{\hbar}^{-t}\Op_{\hbar}(a)\U_{\hbar}^t$ which has principal symbol equals to $a\circ g^t$.
\end{rema}

\subsection{Product of pseudodifferential operators}\label{pa:product}

In the last two paragraphs of this appendix, we fix a smooth partition satisfying the assumptions of paragraph~\ref{s:discretization}. In particular, all the functions $P_j^{\hbar}$ belong to a class of symbol $S^{0,0}_{\overline{\nu}}(T^*M)$ with $0<\overline{\nu}<1/2$. Then, one can verify that the following lemma holds:

\begin{lemm}\label{l:productPDO} Let $\chi_{-j}$ be a cutoff supported in a small neighborhood of $S^*M$ as in paragraph~\ref{ss:cutoff}. There exists $\kappa_0>0$ depending only on $\delta$ (the size of the energy layer), on $\overline{\nu}$ (the parameter for the regularization of the partition) 
and on the choice of the partition such that
$$\forall 0\leq m\leq \kappa_0|\log\hbar|,\ \pi_{\alpha_{m-1}}(m-1)\ldots\pi_{\alpha_1}(1)\pi_{\alpha_0}\Op_{\hbar}(\chi_{-j})$$
is a pseudodifferential operator in $\Psi_{\overline{\nu}_0}^{0,-\infty}(M)$ (where $0<\overline{\nu}_0<1/2$) with principal symbol equal to
$$P_{\alpha_{m-1}}^{\hbar}\circ g^{m-1}\ldots \times P_{\alpha_{1}}^{\hbar}\circ g^{1}\times P_{\alpha_{0}}^{\hbar}\chi_{-j}.$$
\end{lemm}

The proof\footnote{For instance, similar properties on product of pseudodifferential operators were proved in~\cite{Riv10a} (section $7$) in a selfadjoint context. They can be adapted in a nonselfadjoint setting and the situation is even simpler here as we do not try to optimize the parameter $\kappa>0$.} of this lemma relies on the fact that for $\kappa$ small enough, the operators we consider are amenable to semiclassical calculus: composition rules, 
Egorov property (paragraph~\ref{pa:egorov}).

Using this lemma, one also has the following property that we used in the proof of lemma~\ref{l:lowbound}:

\begin{lemm}\label{l:cylsums} Let $\chi_{-j}$ be a cutoff supported in a small neighborhood of $S^*M$ as in paragraph~\ref{ss:cutoff}. There exists $\kappa_0>0$ and $\nu'_0$ depending only on $\delta$ (the size of the energy layer), on $\overline{\nu}$ (the parameter for the regularization of the partition) 
and on the choice of the partition such that for every $0\leq n\leq\kappa_0|\log\hbar|$ and for every subset $W\subset\Sigma_n$,
 $$\left\|\sum_{\gamma\in W}\pi_{\gamma_{n-1}}\ldots\U_{\hbar}\pi_{\gamma_{0}}\U_{\hbar}
\Op_{\hbar}(\chi_{-j})\U_{\hbar}^{-n}\right\|_{L^2(M)\rightarrow L^2(M)}\leq 1+\mathcal{O}(\hbar^{\nu_0'}),$$
where the constant in the remainder is uniform for $W\subset\Sigma_n$.
\end{lemm}

\begin{proof} Let $\gamma$ be an element in $\Sigma_n$. As in lemma~\ref{l:productPDO}, one can verify that for $\kappa_0$ small enough (and uniform in $\gamma$), one has that
$$\forall 0\leq n\leq\kappa_0|\log\hbar|,\ \pi_{\gamma_{n-1}}\ldots\U_{\hbar}\pi_{\gamma_{0}}\U_{\hbar}
\Op_{\hbar}(\chi_{-j})\U_{\hbar}^{-n}$$
is a pseudodifferential operator in $\Psi_{\overline{\nu}'}^{0,-\infty}(M)$ (for some $\overline{\nu}<\overline{\nu}'<1/2$) with principal symbol equal to
$$P_{\gamma_{n-1}}^{\hbar}\ldots \times P_{\gamma_{1}}^{\hbar}\circ g^{1-n}\times P_{\gamma_{0}}^{\hbar}\circ g^{-n}\chi_{-j}.$$
Moreover, there exists $\nu_0>0$ such that, for every $0\leq n\leq\kappa_0|\log\hbar|$, one has
$$\left\|\pi_{\gamma_{n-1}}\ldots\U_{\hbar}\pi_{\gamma_{0}}
\U_{\hbar}\Op_{\hbar}(\chi_{-j})\U_{\hbar}^{-n}-\Op_{\hbar}\left(P_{\gamma_{n-1}}^{\hbar}\ldots \times P_{\gamma_{1}}^{\hbar}\circ g^{1-n}\times P_{\gamma_{0}}^{\hbar}\circ g^{-n}\chi_{-j}\right)\right\|=\mathcal{O}(\hbar^{\nu_0}).$$
One knows that $\sharp W$ is at most equal to $K^n$. Hence, one can verify that for $\kappa>0$ small enough, there exists $\nu_0'>0$ (independent of $\gamma$) such that, for every $0\leq n\leq\kappa|\log\hbar|$,
$$\sum_{\gamma\in W}\pi_{\gamma_{n-1}}\ldots\U_{\hbar}\pi_{\gamma_{0}}\U_{\hbar}
\Op_{\hbar}(\chi_{-j})\U_{\hbar}^{-n}\hspace{6cm}$$
$$\hspace{1cm}=\Op_{\hbar}\left(\sum_{\gamma\in W}P_{\gamma_{n-1}}^{\hbar}\ldots \times P_{\gamma_{1}}^{\hbar}\circ g^{1-n}\times P_{\gamma_{0}}^{\hbar}\circ g^{-n}\chi_{-j}\right)+\mathcal{O}_{L^2(M)\rightarrow L^2(M)}(\hbar^{\nu_0'}).$$
As $(P_i^{\hbar})_{i=1,\ldots, K}$ is a family of nonnegative functions satisfying a property of partition of identity(see paragraph~\ref{s:discretization}), one knows that
$$\left\|\sum_{\gamma\in W}P_{\gamma_{n-1}}^{\hbar}\ldots \times P_{\gamma_{1}}^{\hbar}\circ g^{1-n}\times P_{\gamma_{0}}^{\hbar}\circ g^{-n}\chi_{-j}\right\|_{\infty}\leq 1.$$
Finally, one can verify that $\sum_{\gamma\in W}P_{\gamma_{n-1}}^{\hbar}\ldots \times P_{\gamma_{1}}^{\hbar}\circ g^{1-n}\times P_{\gamma_{0}}^{\hbar}\circ g^{-n}\chi_{-j}$ belongs to some class $S^{-\infty,0}_{\overline{\nu}'}(T^*M)$ for some $0<\overline{\nu}'<1/2$. Thus, one can apply Calder\'on Vaillancourt Theorem (Chapter $5$ in~\cite{EZ} for instance) and it concludes the proof of the lemma. 
\end{proof}

\subsection{Inserting cutoffs functions}

At different stages of the proof, we needed to localize our operators near the energy layer $S^*M$. This can be achieved thanks to to the cutoffs function we introduced in paragraph~\ref{ss:cutoff}. In fact, these cutoffs can be inserted without hurting the quantities we consider as $\Op_{\hbar}(\chi_{-j})\psi_{\hbar}=\psi_{\hbar}+\mathcal{O}(\hbar^{\infty})$ for every $0\leq j\leq k-1$.

For instance, thanks to the properties of our family $(\chi_{-j})_j$, one can prove the following result that we used in lemma~\ref{l:lowbound}:

\begin{lemm}\label{l:cutoff} There exists\footnote{The parameter $\kappa_0$ still depends on on $\delta$ (the size of the energy layer), on $\overline{\nu}$ (the parameter for the regularization of the partition) 
and on the choice of the partition.} $\kappa_0>0$ such that for every cylinder 
$[\gamma]=[\gamma^0,\ldots,\gamma^{k-1}]$ of length $kn$ (where each subcylinder is of length $n\leq[\kappa_0|\log\hbar|]$), one has
$$\left\|\tilde{\Pi}_{\gamma^{k-1}}\Op_{\hbar}(\chi_{0})\ldots\tilde{\Pi}_{\gamma^{1}}\Op_{\hbar}(\chi_{-k+2})
\tilde{\Pi}_{\gamma^{0}}\Op_{\hbar}(\chi_{-k+1})\psi_{\hbar}-\pi_{\gamma^{k-1}_{n-1}}\U_{\hbar}\pi_{\gamma^{k-1}_{n-2}}\U_{\hbar}\ldots\pi_{\gamma^0_{1}}\U_{\hbar}\pi_{\gamma^0_{0}}\psi_{\hbar}\right\|=\mathcal{O}(\hbar^{\infty}),$$
where
$$\tilde{\Pi}_{\gamma^j}=\pi_{\gamma^j_{n-1}}\U_{\hbar}\ldots\pi_{\gamma^j_{1}}\U_{\hbar}\pi_{\gamma^j_{0}}\U_{\hbar},$$
and the remainder is uniform in $\gamma$ and in $\psi_{\hbar}$.
\end{lemm}



\end{document}